\documentclass[a4paper,11pt]{article}
\pdfoutput=1 

\usepackage{jheppub} 

\usepackage[T1]{fontenc} 

\usepackage{tikz}
\usetikzlibrary{calc}
\usetikzlibrary{arrows,decorations.pathmorphing,patterns}
\usepackage{graphicx}

\numberwithin{equation}{section}
\usepackage{version}
\usepackage{mathrsfs}

\title{\boldmath Carroll--Weyl symmetries in flat holography}

\author{Arnaud Delfante$^a$ and Chrysoula Markou$^b$}

\affiliation{$^a$Department of Physics, College of Sciences, and Research Institute of Basic Science,\\ Kyung Hee University, Seoul 02447, Korea}

\affiliation{$^b$Scuola Normale Superiore and INFN,\\ Piazza dei Cavalieri 7, 56126 Pisa, Italy}

\emailAdd{delfante@khu.ac.kr}
\emailAdd{chrysoula.markou@sns.it}

\abstract{The degenerate geometry of a null boundary admits an enlarged local conformal freedom: in two dimensions, a Carrollian coframe allows two independent Weyl rescalings, one modulating and the other preserving the boundary volume form. We investigate whether both transformations can be realized as charged asymptotic symmetries in three-dimensional flat holography. We show that Einstein gravity canonically realizes the volume-modulating Carroll--Weyl transformation, together with local Carroll boosts, but does not admit the volume-preserving rescaling as an additional independent asymptotic symmetry. We trace this obstruction both back to the conformal completion of null infinity and to the absence of two independent commuting semisimple generators in the Poincar\'e algebra. Conformal gravity overcomes this obstruction through bulk dilatations, which supply the missing algebraic direction. We construct an asymptotically flat phase space of conformal gravity in which both types of Carroll--Weyl transformations and local Carroll boosts admit finite, integrable, and generically nonvanishing canonical charges, and derive their centrally extended algebra. Finally, we show that these central extensions arise from boundary anomalies through symplectic descent, yielding a gravitational prediction for Carroll--Weyl anomalies in flat holography.}

\begin{document}

\maketitle
\flushbottom


\section{Introduction} \label{sec. Intro}

Extending the holographic principle beyond anti-de Sitter (AdS) spacetimes to vanishing cosmological constant $\Lambda$ is a major challenge in holography. While the AdS/CFT correspondence provides a concrete realization of the idea that gravitational physics can be encoded in a lower-dimensional quantum theory \cite{Maldacena:1997re,Gubser:1998bc,Witten:1998qj}, considerably less is understood when the conformal boundary becomes null. A holographic description of asymptotically flat spacetimes would not only test which ingredients of AdS/CFT survive beyond its original setting, but could also provide a boundary framework for gravitational scattering and the infrared structure of gravity. In the absence of a complete microscopic formulation, asymptotic symmetries offer a particularly useful guiding principle: canonical transformations of the bulk phase space determine symmetry algebras, currents and Ward identities that any putative boundary description should reproduce \cite{Brown:1986nw,Barnich:2006av,Bagchi:2010zz}. In this work, we follow this principle in three bulk dimensions and ask how much of the intrinsic local symmetry of null infinity can be realized canonically from the bulk.

The geometry of the boundary already suggests that the flat-space problem is qualitatively different from its AdS counterpart. Future null infinity $\mathscr I^+$ is a null hypersurface of the conformally completed spacetime. The metric induced on it is therefore degenerate, with a nowhere-vanishing vector spanning its kernel and selecting a preferred temporal direction \cite{Penrose:1964ge,Ashtekar:2014zsa,Duval:2014uva,Ciambelli:2019lap}. Its intrinsic geometry is Carrollian \cite{Levy-Leblond:1965dsc,Henneaux:1979vn,Duval:2014lpa}, rather than Lorentzian as for the timelike conformal boundary of AdS. Carrollian geometry has moreover emerged as the natural language for a variety of null systems. At black-hole horizons, for instance, the Raychaudhuri and Damour equations \cite{Raychaudhuri:1953yv,Damour:1978cg} can be reorganized as Carrollian conservation equations, leading to a fluid-like description of horizon dynamics \cite{Donnay:2019jiz}. Carrollian structures also arise naturally in flat limits of gravitational boundary data~\cite{Campoleoni:2022wmf,Campoleoni:2023fug} and on the worldsheet of the null string \cite{Bagchi:2013bga}, originally constructed as the tensionless limit of the relativistic string \cite{Schild:1976vq,Karlhede:1986wb,Isberg:1993av}; see also \cite{Ciambelli:2025unn,Ruzziconi:2026bix,Bagchi:2026wcu} for recent reviews.

This perspective suggests asking a simple holographic question. Given the local transformations allowed by the intrinsic geometry of the boundary, which of them can be promoted to genuine asymptotic symmetries of the bulk? The distinction is essential. A transformation may act consistently on the boundary geometric data without being induced by an admissible residual bulk gauge transformation, and a residual transformation may itself remain physically trivial if its canonical generator vanishes. Here we reserve the term \emph{canonical symmetry} for transformations whose associated surface generators are finite, integrable and generically nonvanishing. Enlarging the boundary phase space can promote transformations that would otherwise be treated as gauge redundancies to such improper  gauge symmetries \cite{Regge:1974zd,Barnich:2001jy,Geiller:2021vpg,Campoleoni:2022wmf,Ciambelli:2023ott,Ciambelli:2024vhy,Delfante:2025lxn}. From a holographic viewpoint, each additional charged sector potentially supplies an independent boundary current, enlarges the algebra acting on states and yields further Ward identities. Actually, the remarkable role played by asymptotic symmetries in three-dimensional AdS gravity, where the Brown--Henneaux Virasoro algebra underlies the Cardy derivation of the BTZ entropy \cite{Brown:1986nw,Cardy:1986ie,Banados:1992wn,Strominger:1997eq}, provides an illustration of how much information can be encoded in this boundary symmetry structure.

A distinctive feature of a two-dimensional Carrollian boundary is that its degeneracy allows more local scale freedom than a non-degenerate Lorentzian metric. Locally, the geometry may be described by a Carrollian coframe $(\mathrm{k},\mathrm{n})$, where $\mathrm{k}$ is a clock form and $\mathrm{n}$ spans the one-dimensional spatial cotangent direction, such that the degenerate metric and the volume form read $\mathrm{q}=\mathrm{n}\otimes\mathrm{n}$ and $\mathrm{vol} = \mathrm{k}\wedge\mathrm{n}$, respectively. Since the clock form does not enter the degenerate metric, the temporal and spatial representatives can be rescaled independently \cite{Carinena:1981nq,Duval:2014uva,Hartong:2015xda,Ciambelli:2025unn,Sheikh-Jabbari:2026vqh}. In two dimensions, these two independent local scale freedoms can be combined into two transformations distinguished by their action on the Carrollian volume form. The first rescales $\mathrm{k}$ and $\mathrm{n}$ with the same weight and therefore rescales the Carrollian volume form; the second assigns them opposite weights and leaves the volume invariant. We refer to these transformations as the \emph{volume-modulating} and \emph{volume-preserving Carroll--Weyl rescalings}, respectively. The latter has no analogue as an independent Weyl rescaling of a non-degenerate Lorentzian metric. Together with boundary diffeomorphisms and local Carroll boosts, these transformations constitute the local frame transformations that will be relevant throughout this work.

The existence of this second scaling has recently received particular attention in the context of null strings. As already anticipated, their two-dimensional worldsheets are themselves Carrollian, and recent works have emphasized that their intrinsic geometry admits two independent Weyl-type rescalings rather than the single Weyl transformation of an ordinary tensile string worldsheet \cite{Sheikh-Jabbari:2026vqh,Lindstrom:2026quz,Sheikh-Jabbari:2026cnj,Sheikh-Jabbari:2026tpf,Lindstrom:2026zno}. Gauging the additional, volume-preserving transformation modifies the constraint and ghost structure of the theory and has triggered a renewed analysis of its quantum consistency and anomaly structure \cite{Gustafsson:1994kr,Duary:2026rlo,Duary:2026lmk,Chen:2026cau}. This observation motivates an analogous question at null infinity. Although the usual Carroll--Weyl transformation already has a well-established gravitational realization \cite{Geiller:2021vpg,preprint_Arnaud}, the volume-preserving rescaling has, to the best of our knowledge, not been identified as an independent charged asymptotic symmetry of an asymptotically flat gravitational phase space. Establishing such a realization would be of direct holographic interest: it would enlarge the set of intrinsic boundary transformations that acquire independent canonical currents from the bulk and thereby provide additional structure with which to probe a putative flat-space holographic correspondence.

Our central question is therefore whether the volume-preserving Carroll--Weyl transformation can be realized as a genuine asymptotic symmetry of a three-dimensional asymptotically flat bulk. Three-dimensional Einstein gravity with a vanishing cosmological constant~$\Lambda=0$ provides the natural starting point. Its Chern--Simons formulation, based on the gauge algebra $\mathfrak{iso}(1,2)$, makes the relation between bulk gauge transformations and boundary charges particularly transparent~\cite{Achucarro:1986uwr,Witten:1988hc,Banados:1994tn}. Since the theory carries no local propagating degrees of freedom, its physical content is largely controlled by the choice of boundary conditions. Enlarged asymptotically flat phase spaces are already known to accommodate charged boundary diffeomorphisms, the volume-modulating Carroll--Weyl transformation and local Carroll boosts \cite{Barnich:2006av,Detournay:2016sfv,Afshar:2016kjj,Grumiller:2017sjh,Ciambelli:2018wre,Campoleoni:2018ltl,Adami:2020ugu,Geiller:2021vpg,Adami:2022ktn,Geiller:2022vto,Campoleoni:2022wmf,Campoleoni:2023fug,Geiller:2024amx,Adami:2024rkr,Taghiloo:2024ewx,Delfante:2025lxn,DArcy:2026hgl}. It is thus natural to ask whether a further enlargement of the boundary conditions could also accommodate the second independent rescaling.

We show that an additional independent realization of this transformation is obstructed within asymptotically flat Einstein gravity. The obstruction has both a geometric and an algebraic origin. Geometrically, at fixed physical metric, the usual conformal completion of null infinity correlates the Weyl weights of the degenerate metric and its kernel vector: a change of conformal representative generates the common, volume-modulating Carroll--Weyl transformation, but cannot reproduce the opposite relative scaling required by the volume-preserving one. This provides a gauge-independent obstruction to generating the latter through the conformal completion itself, although an enlarged solution space may still contain boundary data whose independent variations reproduce the corresponding rescaling kinematically. Algebraically, the same distinction becomes transparent in the $\mathfrak{iso}(1,2)$ Chern--Simons description. The diagonal Lorentz generator acts with the same weight on the leading components encoding the temporal and spatial boundary frame, while the relevant translational direction acts nilpotently and generates a Carroll boost. Modulo boundary diffeomorphisms, the Poincar\'e algebra therefore contains no second independent semisimple diagonal generator capable of realizing the required relative rescaling as an additional internal residual symmetry.

Three-dimensional conformal gravity provides precisely such an additional direction. Local Weyl transformations of the physical bulk metric are now genuine gauge symmetries, supplying a scaling freedom independent of the choice of conformal completion. In three dimensions, the theory admits a Chern--Simons formulation with gauge algebra $\mathfrak{so}(3,2)$ \cite{Horne:1988jf}. Its dilatation generator provides the missing semisimple direction, allowing two independent diagonal combinations to act on the leading Carrollian coframe. The appearance of additional boundary currents in conformal gravity is not itself new: a fluctuating bulk Weyl mode was shown in \cite{Afshar:2013bla} to generate an independent boundary current, while more general and near-horizon boundary conditions can retain several Cartan currents \cite{Fuentealba:2020zkf,Lovrekovic:2023xsj,Fuentealba:2024thk}. What has not been established, however, is the identification of these bulk gauge directions with the two intrinsic Carroll--Weyl rescalings of the null-boundary coframe and, in particular, the canonical realization of the volume-preserving transformation in an asymptotically flat phase space.

Guided by this observation, we construct an asymptotically flat phase space of three-dimensional conformal gravity containing an additional boundary field associated with the second diagonal direction. This field produces opposite rescalings of the temporal and spatial components of the Carrollian coframe while leaving its volume form invariant. We show that the corresponding transformation possesses a finite, integrable and generically nonvanishing surface generator, independently of the generator of the volume-modulating Carroll--Weyl transformation. Local Carroll boosts are simultaneously retained as charged symmetries. The two Carroll--Weyl currents form a rank-two centrally extended Abelian sector, whose non-degenerate level matrix is directly inherited from the invariant bilinear form of $\mathfrak{so}(3,2)$, while the boosts participate in mixed Heisenberg-type extensions. The additional conformal direction therefore does more than enlarge the formal gauge algebra of the bulk: once its associated boundary mode is allowed to fluctuate, it becomes a genuinely independent canonical current at null infinity.

The canonical realization of these symmetries is tied to the variational principle and to boundary anomalies. In the Chern--Simons polarization in which the Carroll--Weyl and boost sectors remain charged, the on-shell boundary variation is not strictly invariant under the corresponding residual transformations. We show that this anomalous variation can be removed by a corner improvement of the presymplectic potential. The improvement is not innocuous, however: it simultaneously removes precisely the Carrollian contributions to the surface charges. A strictly invariant variational principle and a non-trivial canonical realization of the enlarged Carrollian symmetry therefore correspond to different choices of symplectic polarization. Moreover, the field-space curvature of the corner term provides a symplectic descent of the boundary anomaly: its antisymmetrized variation reproduces the two-cocycles appearing in the canonical charge algebra, including the matrix-valued Carroll--Weyl central extension and its mixed extensions with the boosts. The anomalous boundary variation, the corner ambiguity and the central extensions are thus different manifestations of the same underlying symplectic structure. We emphasize that the anomaly encountered here is a classical boundary anomaly associated with the gravitational presymplectic structure and should not be directly identified with the quantum Carroll--Weyl anomalies recently studied for tensionless strings. Nevertheless, the parallel appearance of several Carroll--Weyl anomaly cocycles on the two sides suggests a concrete setting in which a possible holographic anomaly matching may eventually be investigated.

The paper is organized as follows. In Section~\ref{sec. C-geom}, we review the intrinsic geometry of a two-dimensional null boundary in terms of its degenerate metric, kernel vector, Carrollian coframe and associated volume form. In Section~\ref{sec. C-sym}, we discuss the local transformations of this structure. Besides boundary diffeomorphisms and Carroll boosts, we identify the two independent Carroll--Weyl rescalings and distinguish them according to whether they modulate or preserve the Carrollian volume. Section~\ref{sec. grav} investigates their holographic realization in three-dimensional asymptotically flat Einstein gravity. We construct a ``covariant Bondi--Weyl'' phase space \cite{Geiller:2021vpg,Campoleoni:2022wmf}, determine its residual transformations, surface charges and asymptotic symmetry algebra, analyze its variational principle and boundary anomalies, and establish the geometric and algebraic obstruction to realizing the volume-preserving Carroll--Weyl transformation. In Section~\ref{sec. conformal-flat}, we turn to the $\mathfrak{so}(3,2)$ Chern--Simons formulation of three-dimensional conformal gravity and construct an enlarged asymptotically flat phase space incorporating the additional boundary degrees of freedom that are sensitive to Carroll--Weyl rescalings. We determine its finite and integrable charges and their centrally extended algebra, relate the Carroll--Weyl level matrix to the invariant bilinear form of $\mathfrak{so}(3,2)$, and analyze the interplay between the variational principle, corner ambiguities and the symplectic descent of the boundary anomaly to the charge-algebra cocycles.


\section{Carroll geometry} \label{sec. C-geom}

We begin with a brief review of Carroll geometry; see, e.g.,~\cite{Henneaux:1979vn,Herfray:2021qmp,Ciambelli:2025unn} for a detailed account. A Carrollian manifold $\Sigma$, with coordinates $x^a$, is equipped with two fundamental structures: a \emph{degenerate} symmetric bilinear form $q_{ab}$ and a nowhere-vanishing vector field $\ell^a$ spanning the kernel of~$q_{ab}$,
\begin{equation} \label{eq. kernel-vector}
    \ell^a q_{ab} = 0 \, .
\end{equation}
For this reason, $\ell:=\ell^a\partial_a$ is referred to as the \emph{kernel vector}. In the following, we specialize the discussion to the holographic setting of null surfaces and, in particular, to null infinity in asymptotically flat (conformal) gravity. We choose coordinates adapted to the null generators such that the kernel vector $\ell$ is aligned with the temporal direction on $\Sigma$.

In the two-dimensional case ($a=0,1$), the intrinsic Carrollian metric $q_{ab}$ has rank one. Since the transverse spatial sector is one-dimensional, the metric can locally be written in terms of a single spatial one-form $n_a$ as
\begin{equation} \label{eq. deg-metric}
    q_{ab} = n_a n_b \, , \qquad \ell^a n_a = 0 \, .
\end{equation}
The one-form $\mathrm{n}:=n_a\mathrm{d}x^a$ defines the spatial component of a Carrollian coframe on $\Sigma$. To complete the coframe, we introduce a \emph{clock form} $\mathrm{k}:=k_a\mathrm{d}x^a$, normalized such that $\mathrm{k}(\ell) = k_a \ell^a = 1$. In contrast with the case of a two-dimensional Lorentzian manifold, the clock form does not enter the intrinsic degenerate metric $q_{ab}$ \eqref{eq. deg-metric}. This distinction will play a crucial role in the following, as it allows for two distinct Weyl rescalings in the context of Carrollian geometry, acting differently on the temporal and spatial components of the coframe. Introducing in addition a spatial vector field $v:=v^a\partial_a$, the coframe $(\mathrm{k},\mathrm{n})$ and its dual frame $(\ell,v)$ obey
\begin{equation} \label{eq. orthonorm}
    \ell^a k_a = 1 \, , \qquad \ell^a n_a = 0 \, , \qquad v^a k_a = 0 \, , \qquad v^a n_a = 1 \, .
\end{equation}
Equivalently, they satisfy the completeness relation
\begin{equation}
    {\delta^a}_b = \ell^a k_b + v^a n_b \, .
\end{equation}

Since the intrinsic metric is degenerate, it does not admit an ordinary notion of inverse. One may nevertheless introduce a \emph{generalized inverse} $q^{ab}$ by requiring it to satisfy
\begin{equation} \label{eq. condgeninv}
    q^{ac} q_{cd} q^{db} = q^{ab} \, .
\end{equation}
Given \eqref{eq. deg-metric} and \eqref{eq. orthonorm}, a natural non-trivial choice in the present two-dimensional setting is
\begin{equation} \label{eq. qvv}
    q^{ab} = v^a v^b \, .
\end{equation}
It follows that
\begin{equation}
    {q^a}_b := q^{ac} q_{cb} = v^a n_b = {\delta^a}_b - \ell^a k_b \, ,
\end{equation}
which is the projector onto the spatial direction transverse to the kernel vector. In particular, the clock form lies in the kernel of the generalized inverse \eqref{eq. qvv},
\begin{equation} \label{eq. gen-inverse}
    q^{ab} k_b = 0 \, .
\end{equation}
From a fiber-bundle perspective, the clock form $\mathrm{k}$ can furthermore be interpreted as defining an \emph{Ehresmann connection} for the Carrollian fibration~\cite{Ciambelli:2019lap}. Indeed, the normalization condition $\mathrm{k}(\ell)=1$ ensures that its kernel defines a horizontal subspace complementary to the vertical direction generated by $\ell$. Accordingly, the tangent bundle decomposes as~$T\Sigma = \mathrm{span}(\ell) \oplus \ker (\mathrm{k})$. In the present two-dimensional setting, $\ker(\mathrm{k})$ is precisely the spatial direction spanned by~$v$. Finally, upon choosing an orientation on $\Sigma$, the corresponding \emph{volume form} is
\begin{equation} \label{eq. vol-form}
    \mathrm{vol}_\Sigma = \mathrm{k} \wedge \mathrm{n} \, .
\end{equation}
%


\section{Carroll symmetries} \label{sec. C-sym}

\subsection{Diffeomorphisms}

The Carrollian geometry introduced in Section~\ref{sec. C-geom} possesses a number of symmetries. First, as for any manifold endowed with a geometric structure, one is free to redefine the coordinates by means of \emph{diffeomorphisms} $y := y^a \partial_a$,
\begin{equation}
    x^a \to x^a + y^a(x) \, .
\end{equation}
As expected, under such transformations the triplet of Carrollian geometric data $(q_{ab},\ell^a,k_a)$ transforms respectively as a rank-two covariant tensor, a vector, and a one-form. Infinitesimally, their transformations are therefore given by the corresponding Lie derivatives:
\begin{equation} \label{eq. C-diffeo}
    \delta_y \! \left(q_{ab},\ell^a,k_a\right) = \mathscr{L}_y \! \left(q_{ab},\ell^a,k_a\right) .
\end{equation}

\subsection{Carroll--Weyl rescalings}

More interestingly, one may observe that the defining relations of the Carrollian geometry admit independent \emph{local rescalings}. The conditions \eqref{eq. kernel-vector} and \eqref{eq. orthonorm} are invariant under a simultaneous rescaling of the kernel vector and its dual clock form,
\begin{equation} \label{eq. 1-scaling}
    \ell^a \to \mathrm{e}^{\omega_1(x)} \ell^a \, , \qquad k_a \to \mathrm{e}^{-\omega_1(x)} k_a \, .
\end{equation}
Independently, the degenerate metric may be locally rescaled according to
\begin{equation} \label{eq. 2-scaling-1}
    q_{ab} \to \mathrm{e}^{2\omega_2(x)} q_{ab} \, .
\end{equation}
In view of \eqref{eq. deg-metric} and \eqref{eq. orthonorm}, this is equivalent to
\begin{equation} \label{eq. 2-scaling-2}
    n_a \to \mathrm{e}^{\omega_2(x)} n_a \, , \qquad v^a \to \mathrm{e}^{-\omega_2(x)} v^a \, .
\end{equation}

The two independent rescalings \eqref{eq. 1-scaling} and
\eqref{eq. 2-scaling-1}--\eqref{eq. 2-scaling-2} can be combined to define two distinct ``Weyl-like'' symmetries of Carrollian geometry. First, choosing $\omega_1=-\omega_2=-\sigma(x)$, one obtains
\begin{equation} \label{eq. fin-1st-CW}
    n_a \to \mathrm{e}^{\sigma} n_a \, , \qquad v^a \to \mathrm{e}^{-\sigma} v^a \, , \qquad k_a \to \mathrm{e}^{\sigma} k_a \, , \qquad \ell^a \to \mathrm{e}^{-\sigma} \ell^a \, .
\end{equation}
Consequently, the degenerate metric \eqref{eq. deg-metric} and its generalized inverse \eqref{eq. qvv} transform with opposite Weyl weights,
\begin{equation} \label{eq. metric-1st-CW}
    q_{ab} \to \mathrm{e}^{2\sigma} q_{ab} \, , \qquad q^{ab} \to \mathrm{e}^{-2\sigma} q^{ab} \, ,
\end{equation}
while the volume form \eqref{eq. vol-form} transforms as
\begin{equation} \label{eq. vol-1st-CW}
    \mathrm{vol}_\Sigma \to \mathrm{e}^{2\sigma}\mathrm{vol}_\Sigma \, .
\end{equation}
This transformation is the Carrollian counterpart of the standard Weyl rescaling of Lorentzian geometry. Infinitesimally, the transformations \eqref{eq. fin-1st-CW} read
\begin{equation} \label{eq. infin-1st-CW}
    \delta_\sigma n_a = \sigma n_a \, , \qquad \delta_\sigma v^a = -\sigma v^a \, , \qquad \delta_\sigma k_a = \sigma k_a \, , \qquad \delta_\sigma \ell^a = -\sigma \ell^a \, ,
\end{equation}
and therefore
\begin{equation}
    \delta_\sigma q_{ab} = 2\sigma q_{ab} \, , \qquad \delta_\sigma q^{ab} = -2\sigma q^{ab} \, , \qquad \delta_\sigma \mathrm{vol}_\Sigma = 2\sigma \,\mathrm{vol}_\Sigma \, .
\end{equation}

Second, one may instead choose $\omega_1 = \omega_2 = \chi(x)$, which leads to
\begin{equation} \label{eq. fin-2nd-CW}
    n_a \to \mathrm{e}^{\chi} n_a \, , \qquad v^a \to \mathrm{e}^{-\chi} v^a \, , \qquad k_a \to \mathrm{e}^{-\chi} k_a \, , \qquad \ell^a \to \mathrm{e}^{\chi} \ell^a \, .
\end{equation}
Hence, at the level of the intrinsic metric, this transformation obeys the same defining rescaling law as an ordinary Weyl transformation (see \eqref{eq. metric-1st-CW}):
\begin{equation} \label{eq. metric-2nd-CW}
    q_{ab} \to \mathrm{e}^{2\chi} q_{ab} \, , \qquad q^{ab} \to \mathrm{e}^{-2\chi} q^{ab} \, .
\end{equation}
However, the $\chi$-action on the temporal and spatial components of the Carrollian dyad is opposite. As a consequence, the volume form is invariant,
\begin{equation} \label{eq. vol-2nd-CW}
    \mathrm{vol}_\Sigma \to \mathrm{vol}_\Sigma \, .
\end{equation}
At the infinitesimal level,
\begin{equation} \label{eq. infin-2nd-CW}
    \delta_\chi n_a = \chi n_a \, , \qquad \delta_\chi v^a = -\chi v^a \, , \qquad \delta_\chi k_a = -\chi k_a \, , \qquad \delta_\chi \ell^a = \chi \ell^a \, ,
\end{equation}
and
\begin{equation}
    \delta_\chi q_{ab} = 2\chi q_{ab} \, , \qquad \delta_\chi q^{ab} = -2\chi q^{ab} \, , \qquad \delta_\chi \mathrm{vol}_\Sigma = 0 \, .
\end{equation}

This thus provides a second, genuinely Carrollian Weyl transformation, with no direct counterpart in non-degenerate Lorentzian geometry. Indeed, in a Lorentzian manifold, the intrinsic metric involves both the temporal and spatial components of the coframe. Consequently, requiring a Weyl rescaling of the full metric ties together the rescalings of these two components, leaving only a single Weyl transformation. In the following, motivated by the different Weyl weights of the volume form under \eqref{eq. vol-1st-CW} and \eqref{eq. vol-2nd-CW}, and borrowing the terminology of~\cite{Sheikh-Jabbari:2026vqh}, we shall refer to the two Carroll--Weyl transformations~\eqref{eq. fin-1st-CW} and~\eqref{eq. fin-2nd-CW} as the \emph{volume-modulating} and \emph{volume-preserving} scalings, respectively.

\subsection{Carroll boosts}

Finally, besides diffeomorphisms and Weyl-like rescalings, Carrollian geometry possesses a local notion of \emph{boosts} \cite{Hartong:2015xda,Ciambelli:2019lap}. This originates from the fact that the choice of the Ehresmann connection is not unique. Indeed, the normalization conditions \eqref{eq. orthonorm}, together with the definition \eqref{eq. gen-inverse} of the generalized inverse, are compatible with a shift of the clock form of the type
\begin{equation}
    k_a \to k_a + \zeta_a \, , \qquad \ell^a \zeta_a = 0 \, .
\end{equation}
Such a shift changes the choice of horizontal distribution, and hence the Ehresmann connection, while leaving the underlying Carrollian structure $(q_{ab},\ell^a)$ unchanged.

In the present two-dimensional setting, any one-form orthogonal to $\ell^a$ is proportional to $n_a$. A \emph{local Carroll boost} is therefore parametrized by a single arbitrary function $\beta(x)$ and acts on the Carrollian coframe as
\begin{equation} \label{eq. fin-CB}
    k_a \to k_a + \beta n_a \, , \qquad n_a \to n_a \, .
\end{equation}
The corresponding transformation of the dual frame is uniquely fixed by requiring the duality relations \eqref{eq. orthonorm} to remain invariant,
\begin{equation}
    \ell^a \to \ell^a \, , \qquad v^a \to v^a - \beta \ell^a \, .
\end{equation}
Consequently, the intrinsic degenerate metric is invariant, whereas its generalized inverse transforms as
\begin{equation}
    q^{ab} \to q'^{ab} = q^{ab} - 2\beta \ell^{(a}v^{b)} + \beta^2 \ell^a\ell^b \, ,
\end{equation}
where parentheses around indices denote symmetrization with unit weight. Nevertheless, together with the transformed clock form $k'_a=k_a+\beta n_a$ \eqref{eq. fin-CB}, the transformed generalized inverse $q'^{ab}$ continues to satisfy $q'^{ab}k'_b=0$ \eqref{eq. gen-inverse}. Moreover, it still obeys $q'^{ac}q_{cd}q'^{db}=q'^{ab}$~\eqref{eq. condgeninv}, consistently reflecting the non-uniqueness of the generalized inverse of a degenerate metric. The volume form is instead also invariant. Infinitesimally, the Carroll boost transformations read
\begin{equation} \label{eq. infin-CB}
    \delta_\beta k_a = \beta n_a \, , \qquad \delta_\beta n_a = 0 \, , \qquad \delta_\beta \ell^a = 0 \, , \qquad \delta_\beta v^a = -\beta \ell^a \, ,
\end{equation}
and therefore
\begin{equation}
    \delta_\beta q_{ab} = 0 \, , \qquad \delta_\beta q^{ab} = -2\beta \ell^{(a}v^{b)} \, , \qquad \delta_\beta \mathrm{vol}_\Sigma = 0 \, .
\end{equation}

\subsection{Summary}

In summary, the local symmetries of the Carrollian geometry can be collected into the following transformations:
\begin{subequations} \label{eq. intrinsic-transf}
    \begin{align}
        \delta_{(y,\sigma,\chi,\beta)} k_a &= \left( \mathscr{L}_y + \sigma - \chi \right) k_a + \beta n_a \, ,\\
        \delta_{(y,\sigma,\chi,\beta)} n_a &= \left( \mathscr{L}_y + \sigma + \chi \right) n_a \, ,\\
        \delta_{(y,\sigma,\chi,\beta)} \ell^a &= \left( \mathscr{L}_y - \sigma + \chi \right) \ell^a \, ,\\
        \delta_{(y,\sigma,\chi,\beta)} v^a &= \left( \mathscr{L}_y - \sigma - \chi \right) v^a - \beta \ell^a \, ,
    \end{align}
\end{subequations}
and
\begin{subequations}
    \begin{align}
        \delta_{(y,\sigma,\chi,\beta)} q_{ab} &= \left( \mathscr{L}_y + 2(\sigma+\chi) \right) q_{ab} \, ,\\
        \delta_{(y,\sigma,\chi,\beta)} q^{ab} &= \left( \mathscr{L}_y - 2(\sigma+\chi) \right) q^{ab} -2\beta \ell^{(a}v^{b)} \, ,\\
        \delta_{(y,\sigma,\chi,\beta)} \mathrm{vol}_\Sigma &= \left( \mathscr{L}_y + 2\sigma \right) \mathrm{vol}_\Sigma \, .
    \end{align}
\end{subequations}

A natural question is then whether these Carrollian symmetries can be realized canonically in flat holography. In the context of three-dimensional asymptotically flat spacetimes, this question has already been extensively investigated for the symmetries generated by $(y,\sigma,\beta)$; see, e.g., \cite{Barnich:2006av,Detournay:2016sfv,Grumiller:2017sjh,Campoleoni:2018ltl,Ciambelli:2020ftk,Adami:2020ugu,Adami:2021nnf,Geiller:2021vpg,Campoleoni:2022wmf,Delfante:2025lxn,DArcy:2026hgl}. To the best of our knowledge, however, the canonical realization of the additional $\chi$-scaling has not yet been explicitly addressed. In the following, we set out to investigate this question.

It is interesting to note that the additional, volume-preserving Carroll--Weyl symmetry has recently played a central role in the reconsideration of the tensionless string~\cite{Gustafsson:1994kr,Sheikh-Jabbari:2026vqh,Lindstrom:2026zno}. While the worldsheet of the tensile string is Lorentzian and admits a single Weyl rescaling, the tensionless limit à la Isberg--Lindstr\"om--Sundborg--Theodoridis (ILST)~\cite{Isberg:1993av} leads to a Carrollian worldsheet \cite{Bagchi:2013bga} and therefore allows the additional relative rescaling discussed above. In particular, the worldsheet metric may be written as a bilinear in terms of a spatial vector $v^a$, as in \eqref{eq. qvv}. However, the ILST formulation does not realize the volume-preserving Weyl transformation as an unrestricted local gauge symmetry, but only retains a residual version in which the Carroll--Weyl parameter is constrained along the null direction. Recent proposals have therefore advocated promoting this transformation to a genuine gauge redundancy with an arbitrary local parameter. This requires a Carroll--Weyl covariantization of the worldsheet theory through the introduction of an additional gauge connection, with the standard ILST formulation recovered after an appropriate gauge fixing~\cite{Sheikh-Jabbari:2026vqh,Sheikh-Jabbari:2026tpf}. This observation has prompted a number of further developments concerning the quantization, anomaly structure and other quantum aspects of Carroll--Weyl-gauged tensionless strings~\cite{Sheikh-Jabbari:2026tpf,Rasulian:2026jvg,Duary:2026rlo,Duary:2026lmk,Chen:2026cau,Lindstrom:2026tua,Sheikh-Jabbari:2026rcr}.

From this perspective, the null string already illustrates an important distinction that will also be central here: although the intrinsic Carrollian geometry admits the additional local rescaling, realizing it with an arbitrary local parameter may require an enlargement of the dynamical framework. It is therefore natural to investigate the same question from the perspective of flat holography and to ask whether the volume-preserving Carroll--Weyl transformation admits a genuine gravitational realization at null infinity. In particular, in analogy with recent arguments in the context of the null string~\cite{Sheikh-Jabbari:2026kwr}, one may ask whether a consistent holographic description of asymptotically flat spacetimes should incorporate the full set of local symmetries allowed by the intrinsic geometry of their null boundary. Attempting such a realization of the additional Carroll--Weyl scaling provides a natural setting in which to address this question.


\section{Asymptotically flat gravity} \label{sec. grav}

We consider \emph{three-dimensional gravity} on an \emph{asymptotically flat} spacetime $\mathcal{M}$, with spacetime coordinates $x^\mu$. For the purposes of the subsequent discussions, it will be more convenient to adopt an algebraic rather than a purely geometric description and to formulate the theory as a \emph{Chern--Simons gauge theory}.

\subsection{Chern--Simons formalism}

Three-dimensional gravity with a vanishing cosmological constant admits the following first-order formulation~\cite{Achucarro:1986uwr,Witten:1988hc}:
\begin{equation} \label{eq. LCS}
    S = \int_{\mathcal{M}} \mathrm{L} \, , \qquad \mathrm{L} = \frac{\kappa}{4\pi} \mathrm{Tr} \! \left(\mathrm{A} \wedge \mathrm{d}\mathrm{A} + \frac{2}{3} \mathrm{A} \wedge \mathrm{A} \wedge \mathrm{A} \right) ,
\end{equation}
where the Chern--Simons level $\kappa$ is fixed in terms of Newton's constant $G$ as $\kappa = \frac{1}{4G}$, and $\mathrm{A}$~is a spacetime one-form taking values in the three-dimensional \emph{Poincar\'e} isometry algebra~$\mathfrak{iso}(1,2)$. We choose the following convenient basis for the latter:
\begin{equation} \label{eq. iso12}
    [J_n,J_m] = (n-m) J_{n+m} \, , \qquad [J_n,P_m] = (n-m) P_{n+m} \, , \qquad [P_n,P_m] = 0 \, ,
\end{equation}
with $n,m\in\{1,0,-1\}$. The generators $J_n$ span the Lorentz subalgebra $\mathfrak{so}(1,2)\simeq\mathfrak{sl}(2,\mathbb{R})$, while the generators $P_n$ generate spacetime translations. In this basis, the Chern--Simons connection decomposes as
\begin{equation}
    \mathrm{A} = \left( {\omega_\mu}^n J_n + {e_\mu}^n P_n \right) \mathrm{d}x^\mu \equiv \left( \omega_\mu + e_\mu \right) \mathrm{d}x^\mu \, ,
\end{equation}
where $e_\mu={e_\mu}^nP_n$ denotes the triad and $\omega_\mu={\omega_\mu}^nJ_n$ its dualized spin connection.

Since the Poincar\'e algebra \eqref{eq. iso12} is non-semisimple, its Killing form is degenerate. The Chern--Simons action \eqref{eq. LCS} is instead defined using the non-degenerate invariant bilinear form
\begin{equation} \label{eq. Triso12}
    \mathrm{Tr}(J_n P_m) = \eta_{nm} = -2 \begin{pmatrix} 0 & 0 & 1 \\ 0 & -\frac{1}{2} & 0 \\ 1 & 0 & 0 \end{pmatrix} ,
\end{equation}
together with
\begin{equation}
    \mathrm{Tr}(J_nJ_m)=0 \, , \qquad \mathrm{Tr}(P_nP_m)=0 \, .
\end{equation}
The spacetime metric on $\mathcal{M}$ is then reconstructed from the triad according to
\begin{equation} \label{eq. gmunu}
    g_{\mu\nu} = {e_\mu}^n \eta_{nm} {e_\nu}^m \, .
\end{equation}

We now introduce \emph{Bondi coordinates} in the bulk, $x^\mu = (r,x^a) = (r,u,\phi)$, where $r$ is the radial coordinate and the asymptotic boundary of interest is future null infinity $\mathscr{I}^+$, reached in the limit $r\to\infty$ at fixed $(u,\phi)$. The coordinates on $\mathscr{I}^+$ are thus $x^a=(u,\phi)$, where $u=t-r$ denotes the retarded time coordinate and $\phi\sim\phi+2\pi$ is the periodic angular coordinate on the celestial circle. In connection with Section~\ref{sec. C-geom}, we henceforth identify the Carrollian manifold introduced there with future null infinity, $\Sigma \equiv \mathscr{I}^+$.

The equations of motion following from \eqref{eq. LCS} simply impose the flatness of the Chern--Simons connection,
\begin{equation} \label{eq. eomCS}
    \mathrm{F} = \mathrm{d}\mathrm{A} + \mathrm{A}\wedge\mathrm{A} = 0 \, .
\end{equation}
They are invariant under finite gauge transformations
\begin{equation} \label{eq. GaugeSym}
    \mathrm{A} \to U^{-1}\mathrm{A} \, U + U^{-1}\mathrm{d}U \, ,
\end{equation}
where $U=\exp(\lambda)\in\mathrm{ISO}(1,2)$ and $\lambda\in\mathfrak{iso}(1,2)$. Infinitesimally, these transformations read
\begin{equation} \label{eq. InfGaugeSym}
    \delta_\lambda\mathrm{A} = \mathrm{d}\lambda + [\mathrm{A},\lambda] \, .
\end{equation}
The associated surface-charge variation is given by \cite{Banados:1994tn}
\begin{equation} \label{eq. CS-charge}
    \delta \mathrm{H}_\lambda = -\frac{\kappa}{2\pi} \int_{\partial\Sigma} \mathrm{Tr}\!\left( \lambda\delta\mathrm{A} \right) .
\end{equation}

It is convenient to exploit the gauge freedom \eqref{eq. GaugeSym} and place the Chern--Simons connection in a \emph{radial gauge},
\begin{equation} \label{eq. radialgauge}
    \mathrm{A}(u,r,\phi) = b^{-1}(r) \left( \mathrm{a}(u,\phi)+\mathrm{d} \right) b(r) \, ,
\end{equation}
so that all explicit radial dependence is carried by the group element $b(r)$, while the reduced connection $\mathrm{a} = a_\mu \mathrm{d}x^\mu$ is independent of~$r$. In particular, by cyclicity of the invariant bilinear form, the surface-charge variation \eqref{eq. CS-charge} can be evaluated entirely in terms of the radially independent data,
\begin{equation}
    \delta \mathrm{H}_\lambda = -\frac{\kappa}{2\pi} \int_{\partial\Sigma} \mathrm{Tr}\!\left( \varepsilon\delta\mathrm{a} \right) ,
\end{equation}
where $\varepsilon=b\lambda b^{-1}$ denotes the corresponding radially independent gauge parameter. Whenever a bulk reconstruction will be needed, we shall adopt the standard Bondi gauge~\cite{Bondi:1960jsa,Sachs:1961zz}, for which the radial group element is chosen as
\begin{equation} \label{eq. bBondi}
    b(r) = \exp\!\left( \frac{r}{2}P_{-1} \right) .
\end{equation}

Following the first-order description of asymptotically flat spacetimes~\cite{Hartong:2015xda,Campoleoni:2022wmf}, we choose the boundary Carrollian coframe to be identified with the leading asymptotic components of the bulk triad. This choice is adapted to the Bondi radial gauge \eqref{eq. bBondi} and is consistent with the conventions of \cite{Riegler:2017fqv}, which we follow in the remainder of this Section.\footnote{Up to the relabeling $L_n\to J_n$ and $M_n\to P_n$.} In this gauge, the bulk metric behaves asymptotically as
\begin{equation} \label{eq. framebulkmetric}
    \mathrm{d}s^2 = g_{\mu\nu} \mathrm{d}x^\mu \mathrm{d}x^\nu = -2 \, \mathrm{d}r \, \mathrm{k} +r^2\mathrm{n}^2 +\mathcal{O}(r) \, ,
\end{equation}
with the boundary metric given by
\begin{equation}
    \mathrm{q} = q_{ab} \mathrm{d}x^a \mathrm{d}x^b = \lim_{r \to \infty} \frac{1}{r^2} \mathrm{d}s^2 = \mathrm{n}^2 \, .
\end{equation}
Comparing with the bulk metric reconstruction \eqref{eq. gmunu}, we accordingly identify
\begin{equation} \label{eq. Carroll-dyad-from-dreibein}
    k_a = {e_a}^{1} \, , \qquad n_a = \lim_{r\to\infty}\frac{1}{r}{e_a}^{0} \, ,
\end{equation}

\subsection{Solution space}

A standard phase space realizing the $\mathrm{BMS}_3$ sector of boundary diffeomorphisms at $\mathscr{I}^+$ is provided by the on-shell \emph{Bondi connections} \cite{Barnich:2006av,Riegler:2017fqv},
\begin{equation} \label{eq. aBondi}
    a_\phi = J_1 + M J_{-1} + N P_{-1} \, , \qquad a_u = P_1+ M P_{-1} \, , \qquad a_r = 0 \, ,
\end{equation}
where
\begin{equation}
    N = L +u M' \, ,
\end{equation}
such that the boundary degrees of freedom are encoded in the Bondi mass aspect $M = M(\phi)$ and the Bondi angular-momentum aspect $L = L(\phi)$, and where a prime denotes a derivative with respect to~$\phi$. Upon radial metric reconstruction using \eqref{eq. gmunu} and \eqref{eq. bBondi}, this corresponds in particular to the standard Bondi--Sachs gauge conditions~\cite{Bondi:1960jsa,Sachs:1961zz}
\begin{equation}
    g_{rr}=g_{r\phi}=0 \, , \qquad g_{ru}=-1 \, , \qquad g_{\phi\phi}=r^2 \, .
\end{equation}

In order to obtain a non-trivial canonical realization of the volume-modulating Carroll--Weyl transformations \eqref{eq. infin-1st-CW} and of the Carroll boosts \eqref{eq. infin-CB}, the standard Bondi phase space must be enlarged. These extensions can be accessed through the \emph{Bondi--Weyl}~(BW) \cite{Geiller:2021vpg} and \emph{covariant Bondi}~(CB) \cite{Campoleoni:2022wmf} gauges, respectively. Here, we revisit these gauges and rederive this statement, while treating them simultaneously for the first time and exploring the consequences of this combined framework.

The Bondi--Weyl extension introduces, in addition to the usual Bondi data, a fluctuating conformal factor $\varphi=\varphi(u,\phi)$ together with a subleading field $H=H(u,\phi)$ associated with an additional freedom in the radial coordinate. Restricting to a conformal boundary frame and setting the Carroll-boost field to zero, the relevant relaxed Bondi--Weyl gauge conditions take the form
\begin{equation}
    g_{rr}=g_{r\phi}=0 \, , \qquad g_{ru}=-\mathrm{e}^\varphi \, , \qquad g_{\phi\phi} = \mathrm{e}^{2\varphi} \left( r - \mathrm{e}^{-\varphi} H \right)^2 \, .
\end{equation}
The covariant Bondi extension further promotes the boundary Carrollian metric to a Carrollian frame by introducing a field $B=B(u,\phi)$ parameterizing local Carroll boosts. In our conventions, the corresponding covariant relaxation of the radial gauge condition reads
\begin{equation}
    g_{rr}=0 \, , \qquad g_{r\phi}=-2 \mathrm{e}^\varphi B \, .
\end{equation}
In the frame $\varphi=H=0$, the angular component of the metric then behaves asymptotically as
\begin{equation}
    g_{\phi\phi} = r^2+8rB\dot B+\mathcal{O}(1) \, ,
\end{equation}
where a dot denotes a derivative with respect to $u$.

In the Chern--Simons formulation, these extensions can be implemented particularly simply by finite $\mathrm{ISO}(1,2)$ gauge transformations. A convenient factorization of the corresponding group element is
\begin{equation}
    g=g_{\mathrm{CB}} g_{\mathrm{BW}} \, ,
\end{equation}
with
\begin{equation} \label{eq. BW-grav-elem}
    g_{\mathrm{BW}} = \exp\!\left( -\frac{H}{2}P_{-1} \right) \exp\!\left( \varphi J_0 \right)
\end{equation}
and
\begin{equation} \label{eq. CB-grav-elem}
    g_{\mathrm{CB}} = \exp\!\left( -\dot B J_{-1} \right) \exp\!\left( 2B P_0-B'P_{-1} \right) .
\end{equation}
The enlarged reduced connection is therefore obtained from the Bondi
representative \eqref{eq. aBondi} as
\begin{equation} \label{eq. CBW-from-Bondi}
    \alpha = g^{-1} \left( \mathrm{a} + \mathrm{d} \right) g \, .
\end{equation}

Equivalently, one has
\begin{subequations} \label{eq. aCBW}
    \begin{align}
        \alpha_\phi &= \mathrm{e}^\varphi \left( J_1+2B P_1 \right) + \left( \varphi'-2\dot B \right)J_0 - H P_0 \nonumber\\
        &\qquad + \mathrm{e}^{-\varphi} \left[ \widehat M J_{-1} + \left( \widehat N-2B\widehat M \right)P_{-1} \right] ,\\
        \alpha_u &= \mathrm{e}^\varphi P_1 + \dot\varphi J_0 + \mathrm{e}^{-\varphi} \left[ \left( \widehat M -\frac{\dot H}{2} +2B\ddot B \right)P_{-1} -\ddot BJ_{-1} \right] ,
    \end{align}
\end{subequations}
where we have introduced the shifted mass and angular-momentum aspects
\begin{equation} \label{eq. shifted-ML}
    \widehat M  = M+\dot B^{2}-\dot B' \, , \qquad \widehat N = N -\frac{H'}{2} -B'' +(H+2B')\dot B \, .
\end{equation}
The associated boundary Carrollian dyad \eqref{eq. Carroll-dyad-from-dreibein} takes the form
\begin{equation} \label{eq. grav-bdy-dyad}
    \mathrm{k} = \mathrm{e}^{\varphi} \left( \mathrm{d}u+2B\,\mathrm{d}\phi \right) , \qquad \mathrm{n} = \mathrm{e}^{\varphi}\mathrm{d}\phi \, ,
\end{equation}
and the corresponding degenerate metric and volume form at future null infinity read
\begin{subequations} \label{eq. qvolgrav}
    \begin{align}
        &\mathrm{q} = q_{ab} \mathrm{d}x^a \mathrm{d}x^b = \mathrm{n}^2 = \mathrm{e}^{2\varphi} \mathrm{d}\phi^2 \, ,\\
        &\mathrm{vol}_{\mathscr I^+} = \mathrm{k}\wedge\mathrm{n} = \mathrm{e}^{2\varphi} \mathrm{d}u\wedge\mathrm{d}\phi \, .
    \end{align}
\end{subequations}
Notice that the clock form $\mathrm{k}$ can also be read off from the $P_1$ component of $\alpha$, while the spatial form $\mathrm{n}$ is encoded in its $J_1$ component. In the present construction, this identification follows directly from the specific form of the solution space \eqref{eq. aCBW} in the gauge \eqref{eq. radialgauge}--\eqref{eq. framebulkmetric}. More intrinsically, however, the same dictionary can be recovered directly at the boundary by gauging the conformal Carroll algebra, building on the conformal-Carroll and Cartan-geometric constructions of, e.g., \cite{Hartong:2015xda,Korovin:2017xqu,Herfray:2021qmp,Figueroa-OFarrill:2022mcy,Bergshoeff:2024ilz}. A detailed derivation of this intrinsic identification will be presented elsewhere.

\subsection{Residual and asymptotic symmetries}

The residual gauge symmetries \eqref{eq. InfGaugeSym} preserving the phase space \eqref{eq. aCBW} are generated by
\begin{align} \label{eq. GravResSym}
    \varepsilon &= \mathrm{e}^\varphi Y J_1 + \left( \sigma-Y'-2Y\dot B \right)J_0 + \mathrm{e}^{-\varphi}\Xi J_{-1} + \mathrm{e}^\varphi \left( f+2BY \right)P_1 \nonumber\\
    &\quad + \mathrm{e}^{-\varphi} \left( \Pi-2B\Xi \right)P_{-1} + \left[ \beta -f' -Y\left(H+2B'\right) -2f\dot B \right]P_0 \, ,
\end{align}
where
\begin{equation} \label{eq. f-def}
    f=T+uY' \, ,
\end{equation}
and
\begin{subequations} \label{eq. XiPi-def}
    \begin{align}
        \Xi &= \frac{1}{2}Y'' +Y'\dot B +Y\left(M+\dot B^{\,2}\right) -\frac{1}{2}\dot\beta \, ,\\
        \Pi &= Y N +Mf +\frac{1}{2}f'' +\frac{1}{2}HY' -\frac{1}{2}h +Y'B' -\frac{1}{2}\beta' \nonumber\\
        &\quad + \left[ f' +Y\left(H+2B'\right) \right]\dot B +f\dot B^{\,2} \, .
    \end{align}
\end{subequations}
The parameters $(Y,T)$ parametrize the residual $\mathrm{BMS}_3$ sector associated with boundary diffeomorphisms. More explicitly, they define the boundary vector field
\begin{equation}
    y = f\,\partial_u + Y\,\partial_\phi \, ,
\end{equation}
with $Y=Y(\phi)$ and $T=T(\phi)$ generating superrotations and supertranslations, respectively. Their action on the Bondi mass and angular-momentum aspects reproduces the standard $\mathrm{BMS}_3$ transformation laws \cite{Barnich:2006av},
\begin{subequations} \label{eq. bms transfo}
    \begin{align}
        \delta_\varepsilon M &= YM' +2Y'M+\frac{1}{2}Y''' \, ,\\
        \delta_\varepsilon L &= YL' +2Y'L+TM' +2T'M+\frac{1}{2}T''' \, .
    \end{align}
\end{subequations}
The additional parameters are defined through
\begin{equation} \label{eq. depsvarphi-depsB}
    \delta_\varepsilon\varphi:=\sigma(u,\phi) \, , \qquad \delta_\varepsilon B:=\frac{1}{2}\beta(u,\phi) \, .
\end{equation}
Accordingly, their action on the boundary dyad \eqref{eq. grav-bdy-dyad} takes the form
\begin{equation}
    \delta_{(\sigma,\beta)}\mathrm{k} =\sigma\,\mathrm{k}+\beta\,\mathrm{n} \, , \qquad \delta_{(\sigma,\beta)}\mathrm{n} =\sigma\,\mathrm{n} \, .
\end{equation}

Actually, in the present parametrization, the superrotations and supertranslations preserve the chosen Carrollian boundary frame, $\delta_{(Y,T)}\mathrm{k}=\delta_{(Y,T)}\mathrm{n}=0$, while acting non-trivially on the Bondi data $(M,L)$. They should therefore be understood as frame-preserving representatives of boundary diffeomorphisms, in which the Lie-derivative action on the Carrollian frame is compensated by Weyl and Carroll-boost transformations. More precisely, denoting by $\hat\sigma$ and $\hat\beta$ the intrinsic Carroll--Weyl and Carroll-boost parameters introduced in Section~\ref{sec. C-sym}, for which $\delta_{(y,\hat{\sigma},\hat{\beta})}\mathrm{k}=\mathscr{L}_y\mathrm{k}+\hat\sigma\mathrm{k}+\hat\beta\mathrm{n}$ and $\delta_{(y,\hat{\sigma},\hat{\beta})}\mathrm{n}=\mathscr{L}_y\mathrm{n}+\hat\sigma\mathrm{n}$, one finds
\begin{equation}
    \sigma=\hat\sigma+f\dot\varphi+Y\varphi'+Y' \, , \qquad \beta=\hat\beta+f'+2f\dot B+2YB' \, .
\end{equation}
When the diffeomorphism sector is switched off, these relations reduce to $\sigma=\hat\sigma$ and $\beta=\hat\beta$, so that $\sigma$ and $\beta$ can then be directly identified with the volume-modulating Carroll--Weyl~\eqref{eq. infin-1st-CW} and Carroll-boost \eqref{eq. infin-CB} parameters, respectively.

The algebra of residual symmetries can be read off as follows. Denoting by
\begin{equation}
    \varepsilon_i = \varepsilon[Y_i,T_i,\sigma_i,\beta_i,h_i] \, , \qquad i=1,2 \, ,
\end{equation}
two independent copies of the residual gauge parameter, the field dependence of the generators requires the use of the modified Lie bracket~\cite{Schwimmer:2008yh,Barnich:2011mi},
\begin{equation} \label{eq. modifiedbracket}
    \varepsilon_{12} := [\varepsilon_1,\varepsilon_2]_\star = [\varepsilon_1,\varepsilon_2] +\delta_{\varepsilon_1}\varepsilon_2 -\delta_{\varepsilon_2}\varepsilon_1 \, , \qquad [\delta_{\varepsilon_1},\delta_{\varepsilon_2}]\alpha = \delta_{\varepsilon_{12}}\alpha \, .
\end{equation}
In terms of the field-independent labels, the resulting composite parameters are
\begin{subequations}
    \begin{align}
        Y_{12} &=Y_2 Y_1' - (1\leftrightarrow2) \, ,\\
        T_{12} &=Y_2 T_1' + T_2 Y_1' - (1\leftrightarrow2) \, ,\\
        \sigma_{12} &= \beta_{12} = h_{12} = 0 \, .
    \end{align}
\end{subequations}
The residual gauge algebra therefore consists of the $\mathfrak{bms}_3$ algebra generated by $(Y,T)$ together with three mutually commuting Abelian sectors generated by $(\sigma,\beta,h)$.

In order to establish a genuine holographic realization of these symmetries within the phase space, one must furthermore verify that the corresponding transformations carry nonvanishing canonical charges. At this stage, the role played by the radially subleading boundary field $H$ and its associated residual gauge parameter
\begin{equation} \label{eq. depsH}
    \delta_\varepsilon H = h
\end{equation}
becomes particularly important. Indeed, the inclusion of this additional boundary degree of freedom allows the Carroll--Weyl transformation generated by $\sigma$ to acquire a non-trivial canonical charge \cite{Geiller:2021vpg}. The integrable surface charge associated with the residual parameter~$\varepsilon$ is
\begin{equation} \label{eq. CBW-charge}
    \mathrm{H}_\varepsilon = \frac{\kappa}{\pi} \int \mathrm{d}\phi  \left[ TM + YL + \beta\dot B - B\dot\beta + \frac{1}{2} \left( \sigma H-h\varphi \right) \right] .
\end{equation}

The different terms in \eqref{eq. CBW-charge} make the canonical realization of the various boundary symmetries manifest. The pair $(Y,T)$ couples to the Bondi angular-momentum and mass aspects $(L,M)$, while the Carroll-boost parameter $\beta$ couples to the boundary frame field~$B$. Finally, the presence of the pair $(H,\varphi)$ renders the volume-modulating Carroll--Weyl transformation generated by $\sigma$ non-trivial in the canonical phase space.

\subsection{Asymptotic symmetry algebra}

The Carrollian symmetries generated by $(y,\sigma,\beta)$ are therefore endowed with nonvanishing canonical generators and are thus genuinely realized holographically in the \emph{covariant Bondi--Weyl} phase space \eqref{eq. CBW-from-Bondi} of three-dimensional asymptotically flat gravity. We now turn to the corresponding charge algebra. We define the Poisson bracket as
\begin{equation} \label{eq. PB}
    \left\{ \mathrm{H}_{\varepsilon_1}, \mathrm{H}_{\varepsilon_2} \right\} := \delta_{\varepsilon_2} \mathrm{H}_{\varepsilon_1} \, .
\end{equation}
For the superrotation and supertranslation sectors, we introduce the usual Fourier modes
\begin{equation}
    \mathcal{J}_n := \mathrm{H}_\varepsilon \left( Y \sim \mathrm{e}^{\mathrm{i}n\phi}
    \right) , \qquad \mathcal{P}_n := \mathrm{H}_\varepsilon \left( T \sim \mathrm{e}^{\mathrm{i}n\phi} \right) , \qquad n \in \mathbb{Z} \, ,
\end{equation}
where all unspecified symmetry parameters are set to zero. Since the Carroll--Weyl and Carroll-boost parameters are arbitrary functions of both $u$ and $\phi$, it is convenient to introduce for these sectors the common double-mode basis
\begin{equation} \label{eq. doublemodebasis}
    F_{pq}(u,\phi) := \mathrm{e}^{\mathrm{i}(p-q)\phi} \mathrm{e}^{\mathrm{i}(p+q)u} \, , \qquad p,q\in\mathbb{R} \, , \quad (p-q) \in \mathbb{Z} \, .
\end{equation}
We accordingly define the volume-modulating Carroll--Weyl modes and their canonically conjugate radial modes as
\begin{equation}
    \mathcal{W}_{pq} := \mathrm{H}_\varepsilon \left( \sigma \sim F_{pq} \right) , \qquad \mathcal{R}_{pq} := \mathrm{i}\, \mathrm{H}_\varepsilon \left( h \sim F_{pq} \right) ,
\end{equation}
together with the Carroll-boost modes
\begin{equation}
    \mathcal{B}_{pq} := \mathrm{H}_\varepsilon \left( \beta=F_{pq} \right) .
\end{equation}
The complete set of nonvanishing Poisson brackets then gives rise to the following $u$-families of infinite-dimensional algebras:
\begin{subequations} \label{eq. full-CBW-algebra}
    \begin{align}
        \mathrm{i} \left\{ \mathcal{J}_n,\mathcal{J}_m \right\} &= (n-m)\mathcal{J}_{n+m} - \frac{c_1}{12} n^3\delta_{n+m,0} \, ,\\
        \mathrm{i} \left\{ \mathcal{J}_n,\mathcal{P}_m \right\} &= (n-m)\mathcal{P}_{n+m} - \frac{c_2}{12} n^3\delta_{n+m,0} \, ,\\
        \mathrm{i} \left\{ \mathcal{W}_{pq},\mathcal{R}_{rs} \right\} &= - \frac{c_2}{12} \mathrm{e}^{2\mathrm{i}(q+s)u}
        \delta_{p+r,q+s} \, ,\\
        \mathrm{i} \left\{ \mathcal{B}_{pq},\mathcal{B}_{rs} \right\} &= -\frac{c_2}{6} (r-q) \mathrm{e}^{2\mathrm{i}(q+s)u}\delta_{p+r,q+s} \, .
    \end{align}
\end{subequations}
The central charges take their standard three-dimensional Einstein-gravity values \cite{Barnich:2006av}
\begin{equation} \label{eq. flat central charge}
    c_1=0 \, , \qquad c_2 = 12\kappa \, .
\end{equation}

The first two brackets span the standard centrally extended $\widehat{\mathfrak{bms}}_3$ algebra. The pair $(\mathcal{W}_{pq},\mathcal{R}_{pq})$ defines a Heisenberg-type extension associated with the volume-modulating Carroll--Weyl symmetry. The Carroll-boost modes instead form a centrally extended local Abelian algebra. Both additional sectors define one-parameter families of charge algebras, parametrized by the retarded time $u$. Their central extensions are, however, qualitatively different: the Carroll--Weyl sector realizes a canonical Heisenberg pairing between the Weyl transformation and its radial partner, whereas the Carroll-boost extension involves a retarded-time derivative and consequently carries the additional factor $(r-q)$.

\subsection{Variational principle and anomalies}

Let us finally discuss the variational principle associated with the enlarged phase space. On shell of the bulk equations of motion, and upon adding the boundary term à la Coussaert--Henneaux--van Driel \cite{Coussaert:1995zp,Campoleoni:2022wmf},
\begin{equation} \label{eq. CHvD}
    S^{(0)}_\mathrm{bdy} = \frac{\kappa}{4\pi} \int \mathrm{d}u \mathrm{d}\phi \, \mathrm{Tr} \! \left( \alpha_u \alpha_\phi \right) ,
\end{equation}
the variation of the total bulk action reduces to
\begin{align}
    \left.\delta (S+S^{(0)}_\mathrm{bdy})\right|_{\mathrm{EOM}} &= \frac{\kappa}{2\pi} \int \mathrm{d}u\mathrm{d}\phi \left[ -2\delta M +\dot H\,\delta\varphi -\dot\varphi\,\delta H -2\delta(\dot B)^2 +4\ddot B\,\delta B \right] ,
\end{align}
up to total derivatives along the periodic celestial circle. The terms which are exact in field space can be removed by adding the boundary functional
\begin{equation}
    S_{\mathrm{bdy}}^{(1)} = \frac{\kappa}{\pi} \int \mathrm{d}u\mathrm{d}\phi \left[ M+(\dot B)^2 \right] .
\end{equation}
The resulting action,
\begin{equation}
    S_{\mathrm{tot}} = S+S_{\mathrm{bdy}}^{(0)}+S_{\mathrm{bdy}}^{(1)}\,,
\end{equation}
has the on-shell variation
\begin{equation} \label{eq. anomalous-variation}
    \left.\delta S_{\mathrm{tot}}\right|_{\mathrm{EOM}} = \frac{\kappa}{2\pi} \int \mathrm{d}u\mathrm{d}\phi \left[ \dot H\,\delta\varphi -\dot\varphi\,\delta H +4\ddot B\,\delta B \right] \equiv \int \mathrm{d}u \mathrm{d}\phi \, \Theta_\mathrm{tot} \, .
\end{equation}
The remaining boundary one-form is not exact in field space and therefore cannot be canceled by an ordinary local boundary counterterm alone. In particular, using \eqref{eq. depsvarphi-depsB} and~\eqref{eq. depsH}, its evaluation on a residual symmetry gives
\begin{equation} \label{eq. anomalous-symmetry-variation}
    \delta_\varepsilon S_{\mathrm{tot}} = \frac{\kappa}{2\pi} \int \mathrm{d}u\mathrm{d}\phi \left( \sigma\dot H -h\dot\varphi +2\beta\ddot B \right).
\end{equation}
This non-invariance may thus be interpreted as the anomalous boundary variation associated with the Carroll--Weyl/radial and Carroll-boost sectors in the present symplectic prescription.

It is nevertheless possible to remove \eqref{eq. anomalous-variation} by exploiting the corner ambiguity \cite{Jacobson:1993vj,Iyer:1994ys,Freidel:2020xyx} of the presymplectic potential $\Theta_\mathrm{tot}$, as discussed in more detail in \cite{Ciambelli:2023ott,Ciambelli:2024vhy,Delfante:2025lxn} in the context of such kinematical charges. Indeed, consider
\begin{equation} \label{eq. gravcornerimprov}
    \frac{2\pi}{\kappa}\vartheta_{\mathrm{corner}} = \varphi\,\delta H -4\dot B\,\delta B \, , \qquad \frac{2\pi}{\kappa}\ell_{\mathrm{bdy}}^{(2)} = 2(\dot B)^2-\varphi\dot H \, .
\end{equation}
One readily verifies that
\begin{equation}
    \Theta_{\mathrm{tot}}+\partial_u\vartheta_{\mathrm{corner}} +\delta\ell_{\mathrm{bdy}}^{(2)} =0 \, .
\end{equation}
Contrary to the addition of an ordinary boundary counterterm, however, the corner improvement modifies the presymplectic form and hence the canonical charges. More precisely,
\begin{equation}
    \frac{2\pi}{\kappa}\delta\vartheta_{\mathrm{corner}} = \delta\varphi\wedge\delta H -4\delta\dot B\wedge\delta B,
\end{equation}
whose contribution to the surface-charge variation is
\begin{equation}
    \Delta(\delta \mathrm{H}_\varepsilon) = -\delta \left[ \frac{\kappa}{2\pi} \int\mathrm{d}\phi \left( \sigma H-h\varphi +2\beta\dot B-2B\dot\beta \right) \right].
\end{equation}
This is precisely minus the $(\sigma,h,\beta)$-contribution to the canonical charge \eqref{eq. CBW-charge}. The improved prescription therefore reduces the latter to
\begin{equation}
    \mathrm{H}_\varepsilon^{\mathrm{improved}} = \frac{\kappa}{\pi} \int\mathrm{d}\phi \left( TM+YL \right),
\end{equation}
thereby rendering the additional non-BMS sectors canonically trivial, including in particular the Carroll--Weyl and Carroll-boost charges. If one wishes instead to retain a non-trivial canonical realization of these Carrollian symmetries, the corner improvement should not be performed, and the anomalous variation \eqref{eq. anomalous-variation} should be kept.

There is an interesting complementary interpretation of the anomalous variation~\eqref{eq. anomalous-symmetry-variation}. Denoting by
\begin{equation}
    \mathcal{A}[\bar{\varepsilon}] := \frac{\kappa}{2\pi} \int \mathrm{d}u\mathrm{d}\phi \left( \sigma\dot H -h\dot\varphi +2\beta\ddot B \right)
\end{equation}
the non-invariant part of the action under a residual transformation
$\bar{\varepsilon}=(\sigma,h,\beta)$, its antisymmetrized second variation reads
\begin{equation} \label{eq. anomaly-descent}
    \delta_{\bar{\varepsilon}_1}\mathcal{A}[\bar{\varepsilon}_2] - \delta_{\bar{\varepsilon}_2}\mathcal{A}[\bar{\varepsilon}_1] = \frac{\kappa}{2\pi} \int \mathrm{d}u\mathrm{d}\phi\, \partial_u \Big[ \sigma_2 h_1-\sigma_1 h_2 +\beta_2\dot\beta_1-\beta_1\dot\beta_2 \Big] \, .
\end{equation}
Thus the Wess--Zumino consistency condition is satisfied up to a contribution localized at the codimension-two corners of the boundary \cite{Wess:1971yu,Chandrasekaran:2020wwn,Freidel:2020xyx}. At fixed retarded time, the corresponding two-cocycle is
\begin{subequations} \label{eq. anomaly-cocycle}
    \begin{align}
        &\delta_{\bar{\varepsilon}_1}\mathcal{A}[\bar{\varepsilon}_2] - \delta_{\bar{\varepsilon}_2}\mathcal{A}[\bar{\varepsilon}_1] = - \int \mathrm{d}u \, \partial_u  K(\bar{\varepsilon}_1,\bar{\varepsilon}_2) \, ,\\
        &K(\bar{\varepsilon}_1,\bar{\varepsilon}_2) = - \frac{\kappa}{2\pi} \int\mathrm{d}\phi \left[ \sigma_2 h_1-\sigma_1 h_2 +\beta_2\dot\beta_1-\beta_1\dot\beta_2 \right] .
    \end{align}
\end{subequations}
This is precisely the cocycle appearing in the canonical charge algebra \cite{Regge:1974zd,Brown:1986ed} for $\bar{\varepsilon}$,
\begin{equation}
    \left\{ \mathrm{H}_{\bar{\varepsilon}_1}, \mathrm{H}_{\bar{\varepsilon}_2} \right\} = \mathrm{H}_{\bar{\varepsilon}_{12}} + K(\bar{\varepsilon}_1,\bar{\varepsilon}_2) \, .
\end{equation}
The first term in~\eqref{eq. anomaly-cocycle} reproduces the central extension between the Carroll--Weyl transformation and its canonically conjugate radial mode,
\begin{equation}
    \mathrm{i} \left\{ \mathcal{W}_{pq},\mathcal{R}_{rs} \right\} = -\frac{c_2}{12}\, \mathrm{e}^{2\mathrm{i}(q+s)u} \delta_{p+r,q+s} \, , \qquad c_2=12\kappa \, ,
\end{equation}
while the second term reproduces the Carroll-boost central extension
\begin{equation}
    \mathrm{i} \left\{ \mathcal{B}_{pq}, \mathcal{B}_{rs} \right\} = -\frac{c_2}{6}(r-q)\, \mathrm{e}^{2\mathrm{i}(q+s)u} \delta_{p+r,q+s} \, .
\end{equation}
The anomalous variation of the action and the central extensions of the canonical algebra are therefore not independent phenomena, as expected from \cite{Delfante:2025lxn}, but are instead related through a symplectic descent. In particular, the corner improvement \eqref{eq. gravcornerimprov} removes the anomalous boundary variation precisely because its field-space curvature cancels the cocycle~\eqref{eq. anomaly-cocycle}. As a consequence, the same improvement also removes the corresponding central extensions and renders the Carroll--Weyl/radial and Carroll-boost generators canonically trivial. Thus, retaining a non-trivial canonical realization of these symmetries and enforcing a strictly invariant variational principle amount to choosing two distinct symplectic polarizations of the same boundary phase space.

Finally, we stress that the anomalous variation discussed here is a classical boundary effect tied to the choice of symplectic potential \cite{Chandrasekaran:2020wwn,Campoleoni:2022wmf} and should not be identified directly with a quantum anomaly of a putative dual theory. Nevertheless, the appearance of Carroll--Weyl and Carroll-boost cocycles suggests an intriguing connection with quantum Carrollian theories and, in particular, with the anomaly structure of tensionless strings~\cite{Bagchi:2021gai,Duary:2026rlo,Duary:2026lmk,Chen:2026cau}. Establishing such a relation would require matching the corresponding cohomology classes and their coefficients, rather than merely comparing their local representatives~\cite{Barnich:2000zw,Campoleoni:2022wmf}.

\subsection{Obstruction to volume-preserving scalings} \label{subsec. no-go-VPCW}

Having established the canonical realization of boundary diffeomorphisms, the volume-modulating Carroll--Weyl symmetry and the local Carroll boosts, it remains to determine whether the second, volume-preserving Carroll--Weyl transformation can also be realized within three-dimensional Einstein gravity. We first address this question from a geometric perspective.

Consider a conformal completion $(\bar{\mathcal M},\bar g_{\mu\nu})$ à la \cite{Penrose:1964ge,Ashtekar:2014zsa} of the asymptotically flat physical spacetime $(\mathcal M,g_{\mu\nu})$, where $\mathcal M$ is the interior of the manifold with boundary $\bar{\mathcal M}$. The physical and unphysical metrics are related by
\begin{equation} \label{eq. bulk Weyl}
    \bar g_{\mu\nu}=\Omega^2 g_{\mu\nu} \, , \qquad \mathscr I^+=\{\Omega=0\} \, ,
\end{equation}
with $\bar g_{\mu\nu}$ extending smoothly to $\mathscr I^+$ and $\mathrm{d}\Omega\neq0$ there. Denoting by $\iota:\mathscr I^+\hookrightarrow\bar{\mathcal M}$ the natural embedding, the pullback $\iota^*$ restricts bulk tensor fields to directions tangent to $\mathscr I^+$. The intrinsic degenerate metric is therefore the metric induced on null infinity,
\begin{equation}
    \mathrm{q} = \iota^*\bar g \, .
\end{equation}
The normal one-form to $\mathscr I^+$ is $\mathrm{d}\Omega$. Raising its index with the unphysical metric defines the vector field
\begin{equation}
    \bar{n} = \bar{n}^\mu \partial_\mu := \bar{g}^{-1}(\mathrm{d}\Omega) \, , \qquad \bar n^\mu := \bar g^{\mu\nu}\partial_\nu\Omega \, .
\end{equation}
Since $\mathscr I^+$ is a null hypersurface, $\bar n^\mu$ is itself tangent to $\mathscr I^+$ there. Its restriction therefore defines the Carrollian kernel vector $\ell$,
\begin{equation}
    \mathrm{d}\iota(\ell) = \left.\bar{n}\right|_{\mathscr I^+} \, .
\end{equation}
In intrinsic coordinates $x^a$ on $\mathscr I^+$, the Carrollian data are thus
\begin{equation} \label{eq. completion-conf}
    q_{ab} = \left(\iota^*\bar g\right)_{ab} \, , \qquad \ell^a = \left.\bar n^a\right|_{\mathscr I^+} \, .
\end{equation}

The conformal completion is not unique. A change of conformal representative is generated by
\begin{equation} \label{eq. OmegaOmegaprime}
    \Omega\rightarrow\Omega' =\mathrm{e}^{\sigma(x)}\Omega \, ,
\end{equation}
which, for fixed physical metric $g_{\mu\nu}$, implies $\bar g_{\mu\nu}\to\bar g'_{\mu\nu} =\mathrm{e}^{2\sigma}\bar g_{\mu\nu}$. It follows that
\begin{equation}
    q_{ab}\rightarrow \mathrm{e}^{2\sigma}q_{ab} \, , \qquad \ell^a\rightarrow \mathrm{e}^{-\sigma}\ell^a \, .
\end{equation}
The universal structure at null infinity is therefore the conformal Carrollian equivalence class \cite{Ashtekar:2014zsa,Herfray:2021qmp}
\begin{equation}
    [\ell^a,q_{ab}] = \left\{ \mathrm{e}^{-\sigma}\ell^a, \mathrm{e}^{2\sigma}q_{ab} \right\}.
\end{equation}
Since $q_{ab}=n_a n_b$ by \eqref{eq. deg-metric}, this implies $\mathrm{n}\to\mathrm{e}^{\sigma}\mathrm{n}$. Upon choosing a clock form $\mathrm{k}$ satisfying $\mathrm{k}(\ell)=1$, the corresponding representative may be chosen to transform as $\mathrm{k}\to\mathrm{e}^{\sigma}\mathrm{k}$, modulo the independent Carroll-boost freedom $\mathrm{k}\to\mathrm{k}+\beta\,\mathrm{n}$. At the level of the Carrollian dyad, this is precisely the volume-modulating Carroll--Weyl transformation \eqref{eq. fin-1st-CW}.

Let us now ask whether the volume-preserving Carroll--Weyl transformation \eqref{eq. fin-2nd-CW} can arise within asymptotically flat Einstein gravity. The argument can be formulated directly in terms of the universal structure of null infinity and is therefore independent of any specific gauge choice. Since the volume-preserving Carroll--Weyl transformation acts as
\begin{equation}
    q_{ab}\rightarrow \mathrm{e}^{2\chi}q_{ab} \, ,
\end{equation}
reproducing its action on the degenerate metric through a change of conformal completion would require $\Omega\to\Omega'=\mathrm{e}^{\chi(x)}\Omega$. However, as we have seen above, any such change of conformal representative necessarily induces
\begin{equation}
    \ell^a\rightarrow \mathrm{e}^{-\chi}\ell^a \, ,
\end{equation}
whereas the volume-preserving Carroll--Weyl transformation \eqref{eq. fin-2nd-CW} requires
\begin{equation}
    \ell^a\rightarrow \mathrm{e}^{\chi}\ell^a \, .
\end{equation}
Equivalently, matching the transformation of $q_{ab}$ requires $\sigma=\chi$, while matching that of $\ell^a$ requires $\sigma=-\chi$. These two conditions are compatible only for the trivial transformation
\begin{equation}
    \chi = 0 \, .
\end{equation}

We therefore arrive at a gauge-independent obstruction to generating the volume-preserving Carroll--Weyl rescaling through the conformal completion alone. At fixed physical Einstein metric, a change of conformal representative necessarily correlates the Weyl weights of the degenerate metric and its kernel vector, and therefore generates only the volume-modulating Carroll--Weyl transformation \eqref{eq. fin-1st-CW}; it cannot reproduce the opposite relative weights required by the volume-preserving transformation \eqref{eq. fin-2nd-CW}. Importantly, this obstruction concerns the realization of the latter, modulo boundary diffeomorphisms, as an additional independent internal residual symmetry, rather than its mere occurrence as a kinematical direction in field space. Indeed, the enlarged three-dimensional Einstein solution space of \cite{Geiller:2025dqe} may contain a priori sufficiently general boundary data to accommodate such a direction with the transformation properties of the volume-preserving Carroll--Weyl rescaling. This is still fully compatible with the present result, since the existence of such a field-space direction does not by itself provide the additional independent internal gauge generator required for its residual realization.

The same obstruction can be understood directly from the asymptotic bulk metric. Schematically, its leading radial behavior takes the form \eqref{eq. framebulkmetric}. A change of radial conformal frame induces the common rescaling of $\mathrm{k}$ and $\mathrm{n}$ associated with the volume-modulating Carroll--Weyl symmetry. By contrast, the volume-preserving transformation \eqref{eq. fin-2nd-CW} requires the two one-forms to carry opposite weights, $\mathrm{k} \rightarrow \mathrm{e}^{-\chi}\mathrm{k}$ and $\mathrm{n} \rightarrow \mathrm{e}^{\chi}\mathrm{n}$, which cannot be generated by a change of conformal representative. Local Lorentz transformations do not circumvent this obstruction: although they can change the choice of frame, they leave the bulk metric, and hence the induced tensor $q_{ab}$, invariant, whereas the transformation above requires $q_{ab}\rightarrow \mathrm{e}^{2\chi}q_{ab}$.

The obstruction also admits a direct algebraic interpretation in the Chern--Simons formulation. As discussed below \eqref{eq. qvolgrav}, the leading components of the connection that encode the Carrollian clock $\mathrm{k}$ and spatial $\mathrm{n}$ structures are carried by the $P_1$ and $J_1$ directions of the Chern--Simons connection, respectively. A local rescaling of the boundary frame must therefore arise from an internal generator acting on these two directions. The common rescaling is generated by $J_0$. Indeed, using
\begin{equation}
    [J_1,J_0]=J_1 \, , \qquad [P_1,J_0]=P_1 \, ,
\end{equation}
one sees that $J_0$ acts diagonally with the same weight on both leading components. It consequently induces the volume-modulating Carroll--Weyl transformation \eqref{eq. fin-1st-CW}. To realize a second, independent local rescaling while preserving this decomposition, one must look for an additional internal direction commuting with $J_0$. In the Poincar\'e algebra,
\begin{equation}
    \operatorname{Cent}_{\mathfrak{iso}(1,2)}(J_0) = \operatorname{span}\{J_0,P_0\}.
\end{equation}
Modulo the already identified $J_0$ transformation, the only remaining candidate is therefore~$P_0$. Its action, however, is qualitatively different. Since
\begin{equation}
    [J_1,P_0]=P_1 \, , \qquad [P_1,P_0]=0 \, ,
\end{equation}
the adjoint action of $P_0$ is nilpotent on the leading frame sector,
\begin{equation}
    (\operatorname{ad}_{P_0})^2=0 \, ,
\end{equation}
and mixes the spatial component into the clock component rather than rescaling the two independently. It thus generates a Carroll boost \eqref{eq. infin-CB}. The homogeneous part of the local frame transformations realized in Einstein gravity consequently takes the triangular form
\begin{equation} \label{eq. Einstein-frame-action}
    \delta_{(\sigma,\beta)} \begin{pmatrix} \mathrm{k}\\[1mm] \mathrm{n} \end{pmatrix} = \left[ \sigma \begin{pmatrix} 1&0\\ 0&1 \end{pmatrix} + \beta \begin{pmatrix} 0&1\\ 0&0 \end{pmatrix} \right] \begin{pmatrix} \mathrm{k}\\[1mm] \mathrm{n} \end{pmatrix}.
\end{equation}
By contrast, the volume-preserving Carroll--Weyl transformation \eqref{eq. fin-2nd-CW} requires a second semisimple direction acting diagonally with opposite weights on the clock and spatial components,
\begin{equation}
    \delta_\chi \begin{pmatrix} \mathrm{k}\\[1mm] \mathrm{n} \end{pmatrix} = \chi \begin{pmatrix} -1&0\\ 0&1 \end{pmatrix} \begin{pmatrix} \mathrm{k}\\[1mm] \mathrm{n} \end{pmatrix}.
\end{equation}
No such independent semisimple action is available within the $\mathfrak{iso}(1,2)$ sector compatible with the leading asymptotic Bondi frame. The Poincar\'e algebra therefore provides one diagonal rescaling and one nilpotent boost, but no second diagonal generator capable of realizing the volume-preserving Carroll--Weyl transformation.

We are thus led to the following algebraic obstruction. Within three-dimensional asymptotically flat Einstein gravity, assuming a non-degenerate bulk triad and the standard $\mathrm{ISO}(1,2)$ Chern--Simons description of null infinity, the residual gauge symmetries can realize the $\mathfrak{bms}_3$ sector associated with boundary diffeomorphisms, the volume-modulating Carroll--Weyl symmetry, and local Carroll boosts. They do not, however, contain an additional independent internal transformation realizing the volume-preserving Carroll--Weyl rescaling.

A useful parallel arises in recent studies of null strings. Promoting the volume-preserving Carroll--Weyl rescaling from the restricted symmetry of the ILST formulation to an unrestricted local gauge symmetry can be achieved by introducing a Carroll--Weyl connection~\cite{Sheikh-Jabbari:2026vqh,Sheikh-Jabbari:2026tpf}. In the curved-background constructions of~\cite{Sheikh-Jabbari:2026kwr}, this gauging appears to be tied to a target-space homothety or, more generally, to an appropriate conformal Killing flow, whose generator can locally be made homothetic by choosing a suitable conformal representative on a patch where it does not vanish. A complementary connection with target-space conformal geometry follows from the interpretation of the gauged model as a reduction of the conformal null string formulated in Dirac's higher-dimensional conformal space~\cite{Lindstrom:2026zno}. Although these constructions are distinct from the gravitational problem considered here, they suggest a structural parallel: realizing the additional Carroll--Weyl rescaling as an unrestricted local symmetry calls for geometric ingredients beyond those retained in the original formulation. In our gravitational setting, the missing ingredient is an independent semisimple bulk gauge direction that generates the volume-preserving rescaling of the boundary coframe. As we now show, three-dimensional conformal gravity supplies precisely this direction through its dilatation generator.


\section{Asymptotically flat conformal gravity} \label{sec. conformal-flat}

The obstruction established in the previous Section identifies the algebraic obstruction to realizing the volume-preserving Carroll--Weyl transformation as a residual, and hence asymptotic, symmetry of three-dimensional Einstein gravity. While the Poincar\'e algebra contains the generators producing a common rescaling of the Carrollian boundary dyad and local Carroll boosts, it lacks an independent semisimple direction assigning opposite weights to its temporal and spatial components. This naturally suggests enlarging the bulk gauge symmetry. We shall argue that three-dimensional conformal gravity provides a natural framework for doing so.

Conformal gravity is governed by the vanishing of the Cotton tensor \cite{Horne:1988jf},
\begin{equation} \label{eq. Cotton}
    C_{\mu\nu} := {\epsilon_\mu}^{\alpha\beta} \, \nabla_\alpha \left( R_{\beta\nu} -\frac{1}{4}R\,g_{\beta\nu} \right) =0 \, .
\end{equation}
In three dimensions, the vanishing of the Cotton tensor is equivalent to local conformal flatness. The equations of motion are covariant under local bulk Weyl transformations
\begin{equation} \label{eq. bulkWeylphysical}
    g_{\mu\nu} \rightarrow \mathrm{e}^{2\rho(x)}g_{\mu\nu} \, ,
\end{equation}
so that Weyl-related metrics describe gauge-equivalent configurations. Importantly, the Cotton equation \eqref{eq. Cotton} does not fix a cosmological constant or select a particular asymptotic geometry. In particular,
\begin{equation}
    R_{\mu\nu}=0 \qquad\Longrightarrow\qquad C_{\mu\nu}=0 \, ,
\end{equation}
so that every vacuum asymptotically flat Einstein solution in three dimensions is also a solution of conformal gravity. The same $\mathrm{SO}(3,2)$ Chern--Simons theory thus admits both asymptotically AdS and asymptotically flat sectors, which are therefore distinguished by their boundary conditions rather than by the bulk gauge algebra \cite{Horne:1988jf,Afshar:2013bla}. In the following, by \emph{asymptotically flat conformal gravity} we shall mean the sector whose conformal class contains an asymptotically flat representative admitting a smooth future null infinity. Bulk Weyl transformations whose Weyl factor extends smoothly and remains nonvanishing at~$\mathscr I^+$ preserve the null character of the conformal boundary and map its intrinsic Carrollian structure into a conformally related one.

Actually, already at the level of the Penrose conformal completion, one can see geometrically how conformal gravity evades the obstruction encountered in Einstein gravity. Besides changing the defining function, as we have seen above \eqref{eq. bulkWeylphysical}, we may now also perform an independent Weyl transformation of the physical bulk metric, together with a change of conformal representative \eqref{eq. OmegaOmegaprime}
\begin{equation}
    \Omega\rightarrow \mathrm{e}^{\tau(x)}\Omega \, .
\end{equation}
The unphysical metric \eqref{eq. bulk Weyl} then transforms as
\begin{equation}
    \bar g_{\mu\nu} \rightarrow \bar g'_{\mu\nu} = \mathrm{e}^{2(\tau+\rho)}\bar g_{\mu\nu} \, ,
\end{equation}
and the induced Carrollian data \eqref{eq. completion-conf} at $\mathscr{I}^+$ consequently obey
\begin{equation}
    q_{ab}\rightarrow \mathrm{e}^{2(\tau+\rho)}q_{ab} \, , \qquad \ell^a\rightarrow \mathrm{e}^{-\tau-2\rho}\ell^a \, .
\end{equation}
The two independent Carroll--Weyl transformations \eqref{eq. fin-1st-CW} and \eqref{eq. fin-2nd-CW}, parametrized by $(\sigma,\chi)$, are therefore reproduced by choosing
\begin{equation}
    \rho=-2\chi \, ,\qquad \tau=\sigma+3\chi \, .
\end{equation}

In particular, it means that a pure volume-preserving transformation \eqref{eq. fin-2nd-CW} requires both $g_{\mu\nu}\rightarrow\mathrm{e}^{-4\chi}g_{\mu\nu}$ and $\Omega\rightarrow\mathrm{e}^{3\chi}\Omega$. For an arbitrary local parameter $\chi(x)$, the first transformation is not a gauge symmetry of Einstein gravity, since local Weyl rescalings of the physical metric do not in general preserve the Einstein equations. By contrast, it is a genuine gauge symmetry of three-dimensional conformal gravity and maps solutions of the Cotton equation into solutions. The additional Weyl freedom \eqref{eq. bulkWeylphysical} of the physical bulk metric therefore supplies precisely the second independent local rescaling that is absent in the conformal completion of a fixed Einstein metric. We now demonstrate this mechanism algebraically in the Chern--Simons formulation and establish its explicit canonical realization at null infinity.

\subsection{Chern--Simons formalism}

Three-dimensional conformal gravity admits a Chern--Simons formulation based on the conformal algebra $\mathfrak{so}(3,2)$ \cite{Horne:1988jf}. The action retains the generic form \eqref{eq. LCS},\footnote{Note that, unlike in Einstein gravity, the Chern--Simons level $\kappa$ of conformal gravity is not fixed in terms of Newton's constant, but instead constitutes an independent dimensionless coupling.} but the gauge connection is now valued in $\mathfrak{so}(3,2)$ and decomposes as
\begin{equation} \label{eq. conformal-connection}
    \mathrm{A} = \left( {\omega_\mu}^n J_n + {e_\mu}^n P_n + {f_\mu}^n K_n + d_\mu D \right)\mathrm{d}x^\mu \, .
\end{equation}
Besides the Lorentz generators $J_n$ and translations $P_n$ spanning
the gravitational subalgebra~$\mathfrak{iso}(1,2)$, the conformal algebra contains the special conformal generators $K_n$ and the dilatation generator $D$. We employ the modal basis $n,m\in\{-1,0,1\}$ adapted to the~$\mathfrak{iso}(1,2)$ embedding \eqref{eq. iso12} used in the previous Section. This basis is particularly well suited to the asymptotically flat sector of conformal gravity, as it keeps manifest the Poincar\'e subalgebra underlying three-dimensional flat gravity while embedding it directly into the full conformal algebra. It also matches the wedge generators of $\mathfrak{bms}_3$, and in this basis $\mathfrak{so}(3,2)$ coincides with the wedge algebra of the conformal extension of $\mathfrak{bms}_3$~\cite{Fuentealba:2020zkf}. Its nonvanishing commutation relations are
\begin{subequations} \label{eq. so32-algebra}
    \begin{align}
        [J_n,J_m] &= (n-m)J_{n+m} \, ,\\
        [J_n,P_m] &= (n-m)P_{n+m} \, ,\\
        [J_n,K_m] &= (n-m)K_{n+m} \, ,\\
        [P_n,D] &= P_n \, ,\\
        [K_n,D] &= -K_n \, ,\\
        [P_n,K_m] &= -2(n-m)J_{n+m} - 2(n^2-nm+m^2-1)D \, .
    \end{align}
\end{subequations}

Since $\mathfrak{so}(3,2)$ is simple, its invariant symmetric bilinear form is unique up to an overall normalization, in contrast with the non-semisimple Poincar\'e algebra $\mathfrak{iso}(1,2)$. Choosing this normalization such that $\mathrm{Tr}(J_nJ_m)=\eta_{nm}$, its nonvanishing components in the basis~\eqref{eq. so32-algebra} are
\begin{equation} \label{eq. so32-bilinear}
    \mathrm{Tr}(J_nJ_m) = \eta_{nm} \, , \qquad \mathrm{Tr}(P_nK_m) = -2\eta_{nm} \, , \qquad \mathrm{Tr}(DD) = 1 \, .
\end{equation}
All other pairings vanish. In particular, $\mathrm{Tr}(J_nP_m)=0$, in contrast with the invariant bilinear form \eqref{eq. Triso12} used in the $\mathfrak{iso}(1,2)$ Chern--Simons formulation of Einstein gravity, for which the Lorentz and translation generators are paired. Consequently, although Einstein solutions can be embedded into the solution space of conformal gravity at the level of the bulk equations of motion, this embedding does not preserve the symplectic structure. The two theories are equipped with different invariant bilinear forms and hence with different presymplectic structures. Their canonical charges must therefore be computed independently.

The equations of motion $\mathrm{F}[\mathrm{A}] = \mathrm{d}\mathrm{A} + \mathrm{A} \wedge \mathrm{A} = 0$ once again impose the flatness of the full conformal connection, as in \eqref{eq. eomCS}. In the presence of the dilatation gauge field $d$, the $P_n$ component of the flatness condition imposes the vanishing of the Weyl-covariant torsion. For an invertible triad, special conformal transformations can be used to fix the gauge $d=0$. In this gauge, the flatness equations impose the ordinary torsion constraint, determine the special-conformal field $f^n$ in terms of the Schouten tensor, and reduce the remaining dynamical equation to the vanishing of the Cotton tensor \eqref{eq. Cotton} \cite{Horne:1988jf}. The spacetime metric is reconstructed from the triad \eqref{eq. gmunu} as before, while a gauge transformation generated by~$D$ acts as an ordinary Weyl rescaling,
\begin{equation}
    \delta_\rho \mathrm{e}^n=\rho \, \mathrm{e}^n \qquad\Rightarrow\qquad \delta_\rho g_{\mu\nu} = 2\rho \, g_{\mu\nu} \, .
\end{equation}
The existence of an asymptotically flat conformal sector is particularly transparent in this formulation. Let
\begin{equation}
    \mathrm{A}_{\mathrm{E}} = \omega_{\mathrm{E}}^nJ_n +e_{\mathrm{E}}^nP_n
\end{equation}
be an asymptotically flat Einstein connection, regarded as an $\mathfrak{so}(3,2)$ connection through the Poincar\'e embedding. Starting from this flat connection, consider the finite dilatation gauge transformation
\begin{align} \label{eq. conformal-flat-embedding}
    \mathrm{A}_{\mathrm{C}} &= \mathrm{e}^{-\psi(x) D} \left( \mathrm{A}_{\mathrm{E}}+\mathrm{d} \right) \mathrm{e}^{\psi(x) D} \nonumber\\
    &= \omega_{\mathrm{E}}^nJ_n +\mathrm{e}^{\psi}e_{\mathrm{E}}^nP_n +\mathrm{d}\psi\,D \, .
\end{align}
Being gauge-related to $\mathrm{A}_{\mathrm{E}}$, the connection $\mathrm{A}_{\mathrm{C}}$ still satisfies $\mathrm{F}=0$. The corresponding metric is
\begin{equation}
    g_{\mu\nu} = \mathrm{e}^{2\psi} g_{\mu\nu}^{\mathrm{E}} \, ,
\end{equation}
which is precisely the bulk Weyl transformation \eqref{eq. bulkWeylphysical}. Provided that the Weyl factor extends smoothly and remains nonvanishing at $\mathscr I^+$, the transformed metric belongs to the same asymptotically flat conformal sector and admits a smooth null conformal boundary.

Notice that the representative \eqref{eq. conformal-flat-embedding} has $f^n=0$ and $d=\mathrm{d}\psi$. The condition $f^n=0$ should not be regarded as an invariant truncation of the full $\mathfrak{so}(3,2)$ gauge theory for asymptotically flat spacetimes, since generic conformal transformations involving $K_n$ do not preserve it. However, the generators~$\{J_n,P_n,D\}$ form a closed subalgebra that does preserve this restriction. This sector already contains the transformations required for our purposes: the two independent Carroll--Weyl rescalings and the local Carroll boosts. We shall therefore restrict our analysis to this sector rather than attempting to realize the full conformal extension of $\mathfrak{bms}_3$; see, e.g.,~\cite{Fuentealba:2020zkf,Fuentealba:2024thk,Grumiller:2026kgx} for related analyses of conformal BMS symmetries and their extensions. We will nevertheless discuss below how the solution space considered here relates to those studied in these works.

The key algebraic feature of the conformal extension is that the enlarged gauge algebra~\eqref{eq. so32-algebra} now contains two independent commuting directions acting diagonally on the leading~$(P_1,J_1)$ doublet. A convenient basis for them is
\begin{equation} \label{eq. TpTm}
    T_+ := J_0 \, , \qquad T_- := 2D-J_0 \, .
\end{equation}
Indeed,
\begin{subequations}
    \begin{align}
        [P_1,T_+] &= P_1 \, , &[J_1,T_+] &= J_1 \, ,\\
        [P_1,T_-] &= P_1 \, , &[J_1,T_-] &= -J_1 \, .
    \end{align}
\end{subequations}
Since the leading $P_1$ and $J_1$ components of the Chern--Simons connection encode the Carrollian clock and spatial one-forms, respectively (see the discussion below \eqref{eq. qvolgrav}), the gauge parameter
\begin{equation}
    \varepsilon_{\mathrm{diag}} = \sigma(x) T_+ - \chi(x) T_-
\end{equation}
generates, through the homogeneous part of the gauge transformation,
\begin{equation}
    \delta_{\mathrm{diag}}\mathrm{k} = (\sigma-\chi)\mathrm{k} \, , \qquad \delta_{\mathrm{diag}}\mathrm{n} = (\sigma+\chi)\mathrm{n} \, .
\end{equation}
These are precisely the infinitesimal form of the two independent Carroll--Weyl rescalings introduced in Section~\ref{sec. C-sym}. This algebraic observation alone does not yet establish their canonical realization. For local parameters $\sigma(x)$ and $\chi(x)$, the leading diagonal gauge parameter must be supplemented by the additional components required to preserve the full asymptotic connection, and the associated surface charges must be shown to be finite, integrable, and nonvanishing. This is the purpose of the following analysis.

\subsection{Solution space}

We now construct an asymptotically flat solution space of conformal gravity which contains, as a subsector at the level of field configurations, the covariant Bondi--Weyl family introduced in Section~\ref{sec. grav}, and which accommodates the two independent Carroll--Weyl rescalings~\eqref{eq. infin-1st-CW}--\eqref{eq. infin-2nd-CW}. The construction below constitutes a minimal ansatz tailored to this purpose. As we shall see, it is nevertheless sufficient to realize canonically the Carrollian symmetries of interest. Starting from the Bondi connection~\eqref{eq. aBondi}, we consider the finite~$\mathrm{SO}(3,2)$ gauge transformation
\begin{equation} \label{eq. conformal-group-element}
    g = g_{\mathrm{CB}}g_{\mathrm{BW}} \, ,
\end{equation}
where the covariant Bondi group element retains the form \eqref{eq. CB-grav-elem}, while the Bondi--Weyl element \eqref{eq. BW-grav-elem} is extended to
\begin{equation}
    g_{\mathrm{BW}} = \exp\!\left( -\frac{H}{2}P_{-1} \right) \exp\!\left( \varphi T_+-\varpi T_- \right) .
\end{equation}
In addition to the boundary fields $B=B(u,\phi)$, $\varphi=\varphi(u,\phi)$ and $H=H(u,\phi)$ inherited from the Einstein-gravity solution space, we introduce a new independent field $\varpi=\varpi(u,\phi)$ along the second diagonal direction $T_-=2D-J_0$ of \eqref{eq. TpTm}, made available by the conformal extension of the bulk gauge algebra. As will become explicit below, $\varpi$ encodes the independent volume-preserving Carroll--Weyl degree of freedom. Notice that, since $[T_+,T_-]=0$, the diagonal part of the above group element can equivalently be factorized as
\begin{equation} \label{eq. conformal-diagonal-element}
    \exp\!\left( \varphi T_+-\varpi T_- \right) = \exp\!\left( (\varphi+\varpi)J_0 \right) \exp\!\left( -2\varpi D \right) .
\end{equation}

The conformal covariant Bondi--Weyl connection is then obtained from \eqref{eq. CBW-from-Bondi}, and equivalently reads
\begin{subequations} \label{eq. conf-aCBW}
    \begin{align}
        \alpha_\phi &= \mathrm{e}^{\varphi+\varpi} J_1 + 2\mathrm{e}^{\varphi-\varpi} B P_1 + \left( \varphi'-2\dot B \right)T_+ - \mathrm{e}^{-2\varpi}H P_0 \nonumber\\
        &\quad + \mathrm{e}^{-\varphi-\varpi} \left[ \widehat M J_{-1} + \mathrm{e}^{-2\varpi} \left( \widehat N-2B\widehat M \right)P_{-1} \right] - \varpi' T_- \, ,\\
        \alpha_u &= \mathrm{e}^{-\varphi-\varpi} \left[ \mathrm{e}^{-2\varpi} \left( \widehat M -\frac{\dot H}{2} + 2 B \ddot B \right)P_{-1} - \ddot B J_{-1} \right] \nonumber\\
        &\quad + \mathrm{e}^{\varphi-\varpi} P_1 + \dot\varphi T_+ - \dot\varpi T_- \, .
    \end{align}
\end{subequations}
Here we have again used the shifted Bondi aspects introduced in \eqref{eq. shifted-ML}. Notice that the introduction of $\varpi$ does not induce any additional shift of $\widehat M$ or $\widehat N$. Rather, the new field acts on the corresponding components of the connection solely through the diagonal adjoint weights generated by $T_-$. We furthermore remain in the radial gauge \eqref{eq. radialgauge}, with \eqref{eq. bBondi} as the radial group element. The associated boundary Carrollian dyad takes the form
\begin{equation} \label{eq. conformal-boundary-dyad}
    \mathrm{k} = \mathrm{e}^{\varphi-\varpi} \left( \mathrm{d}u+2B\,\mathrm{d}\phi \right) , \qquad \mathrm{n} = \mathrm{e}^{\varphi+\varpi} \mathrm{d}\phi \, .
\end{equation}
The corresponding degenerate metric and volume form at future null infinity are therefore
\begin{subequations} \label{eq. conformal-boundary-volume}
    \begin{align}
        &\mathrm{q} = q_{ab} \mathrm{d}x^a \mathrm{d}x^b = \mathrm{n}^2 = \mathrm{e}^{2(\varphi+\varpi)} \mathrm{d}\phi^2 \, ,\\
        &\mathrm{vol}_{\mathscr I^+} = \mathrm{k}\wedge\mathrm{n} = \mathrm{e}^{2\varphi} \mathrm{d}u\wedge\mathrm{d}\phi \, .
    \end{align}
\end{subequations}

The geometric roles of the two conformal fields are thus manifest. The field $\varphi$ produces a common rescaling of the two legs of the Carrollian dyad and consequently controls the local Carrollian volume, as in Einstein gravity, whereas $\varpi$ rescales the temporal and spatial legs with opposite weights while leaving the volume form invariant. They therefore encode, respectively, the volume-modulating \eqref{eq. fin-1st-CW} and volume-preserving Carroll--Weyl \eqref{eq. fin-2nd-CW} degrees of freedom. This identification becomes even more transparent by examining the residual gauge transformations.

\subsection{Residual and asymptotic symmetries}

The residual gauge transformations preserving the conformal phase space are generated by
\begin{align} \label{eq. conf-residual-parameter}
    \varepsilon &= \mathrm{e}^{\varphi+\varpi} Y J_1 + \left( \sigma-Y'-2Y\dot B \right) T_+ - \chi T_- + \mathrm{e}^{-\varphi-\varpi} \Xi J_{-1} \nonumber\\
    &\quad + \mathrm{e}^{\varphi-\varpi} \left( f+2BY \right) P_1 + \mathrm{e}^{-\varphi-3\varpi} \left( \Pi-2B\Xi \right) P_{-1} \nonumber\\
    &\quad + \mathrm{e}^{-2\varpi} \left[ \beta-f'-Y\left(H+2B'\right)-2f\dot B \right] P_0 \, ,
\end{align}
where we have again used the parameter $f$ introduced in \eqref{eq. f-def} and the functions $(\Xi,\Pi)$ defined in \eqref{eq. XiPi-def}. In particular, the introduction of the new conformal field $\varpi$ does not modify the functional form of $(f,\Xi,\Pi)$ inherited from the Einstein-gravity solution space; it only dresses the various components of the residual gauge parameter with the appropriate~$T_-$ weights. The independent variations of the boundary fields are parametrized by
\begin{equation}
    \delta_\varepsilon\varphi := \sigma(u,\phi) \, , \qquad \delta_\varepsilon\varpi := \chi(u,\phi) \, , \qquad \delta_\varepsilon B := \frac{1}{2}\beta(u,\phi) \, , \qquad \delta_\varepsilon H := h(u,\phi) \, .
\end{equation}
The Bondi mass and angular-momentum aspects $(M,L)$ continue to transform according to the standard $\mathfrak{bms}_3$ laws \eqref{eq. bms transfo}.

The geometric interpretation of the new residual parameters can be read directly from their action on the boundary Carrollian dyad \eqref{eq. conformal-boundary-dyad}. Switching off the boundary diffeomorphism sector, $Y=T=0$, one obtains
\begin{equation}
    \delta_{(\sigma,\chi,\beta)}\mathrm{k} = \left(\sigma-\chi\right)\mathrm{k} +\mathrm{e}^{-2\varpi}\beta\,\mathrm{n} \, , \qquad \delta_{(\sigma,\chi,\beta)}\mathrm{n} = \left(\sigma+\chi\right)\mathrm{n} \, .
\end{equation}
Thus, $\sigma$ generates the volume-modulating Carroll--Weyl rescaling, while $\chi$ generates the independent volume-preserving Carroll--Weyl rescaling. The parameter $\beta$ generates a Carroll boost, although it differs from the intrinsic boost parameter because of the $\varpi$-dependent normalization of the Carrollian frame. Denoting the latter by $\hat{\beta}$, one finds, for vanishing boundary diffeomorphisms,
\begin{equation} \label{eq. hatbetaconf}
    \hat{\beta} = \mathrm{e}^{-2\varpi}\beta \, .
\end{equation}

Actually, as in Einstein gravity, the parameters $(Y,T)$ provide frame-preserving representatives of the residual boundary diffeomorphisms. More precisely, setting $\sigma=\chi=\beta=0$ while keeping $(Y,T)$ nonvanishing leaves the chosen Carrollian frame invariant,
\begin{equation}
    \delta_{(Y,T)}\mathrm{k} = \delta_{(Y,T)}\mathrm{n} = 0 \, ,
\end{equation}
while acting non-trivially on the Bondi data $(M,L)$. This should not be confused with the action of a bare boundary diffeomorphism on the Carrollian frame. Rather, the Lie-derivative action generated by
\begin{equation}
    y = f\,\partial_u+Y\,\partial_\phi \, , \qquad f=T+uY' \, ,
\end{equation}
is compensated by suitable Carroll--Weyl and Carroll-boost transformations.

Indeed, the Lie derivatives of the dyad are
\begin{subequations}
    \begin{align}
        \mathscr{L}_y\mathrm{n} &= \left[ f(\dot\varphi+\dot\varpi) +Y(\varphi'+\varpi') +Y' \right]\mathrm{n} \, ,\\
        \mathscr{L}_y\mathrm{k} &= \left[ f(\dot\varphi-\dot\varpi) +Y(\varphi'-\varpi') +Y' \right]\mathrm{k} \nonumber\\
        &\quad +\mathrm{e}^{-2\varpi} \left( f'+2f\dot B+2YB' \right)\mathrm{n} \, .
\end{align}
\end{subequations}
Denoting by $(\hat{\sigma},\hat{\chi},\hat{\beta})$ the intrinsic Carroll--Weyl and Carroll-boost parameters introduced in Section~\ref{sec. C-sym}, such that
\begin{equation}
    \delta\mathrm{k} = \mathscr{L}_y\mathrm{k} +(\hat{\sigma}-\hat{\chi})\mathrm{k} +\hat{\beta}\,\mathrm{n} \, , \qquad \delta\mathrm{n} = \mathscr{L}_y\mathrm{n} +(\hat{\sigma}+\hat{\chi})\mathrm{n} \, ,
\end{equation}
the residual parameters are related to the intrinsic geometric ones by
\begin{subequations}
    \begin{align}
        \sigma &= \hat{\sigma} +f\dot\varphi +Y\varphi' +Y' \, ,\\
        \chi &= \hat{\chi} +f\dot\varpi +Y\varpi' \, ,\\
        \beta &= \mathrm{e}^{2\varpi} \hat{\beta} +f' +2f\dot B +2YB' \, .
    \end{align}
\end{subequations}
When the boundary diffeomorphism sector is switched off, these relations reduce to $\sigma=\hat{\sigma}$, $\chi=\hat{\chi}$ and $\hat{\beta}=\mathrm{e}^{-2\varpi}\beta$, as in \eqref{eq. hatbetaconf}.

The algebra of residual symmetries can be determined using the modified bracket~\eqref{eq. modifiedbracket}, which is required by the field dependence of the gauge parameters \eqref{eq. conf-residual-parameter}. Let
\begin{equation}
    \varepsilon_i = \varepsilon[Y_i,T_i,\sigma_i,\chi_i,\beta_i,h_i] \, , \qquad i=1,2 \, ,
\end{equation}
denote two independent residual gauge parameters. In terms of these
field-independent labels, their modified bracket closes according to
\begin{subequations}
    \begin{align}
        Y_{12} &= Y_2 Y_1' - (1 \leftrightarrow 2) \, ,\\
        T_{12} &= Y_2 T_1' + T_2 Y_1' - (1 \leftrightarrow 2) \, ,\\
        \sigma_{12} &= \chi_{12} = \beta_{12} = h_{12}=0 \, .
    \end{align}
\end{subequations}
In this field-dependent, compensated parametrization, the residual gauge algebra therefore consists of the $\mathfrak{bms}_3$ algebra generated by $(Y,T)$ together with four mutually commuting Abelian sectors generated by $(\sigma,\chi,\beta,h)$. It is worth stressing that this manifest direct-sum structure is a property of the compensated residual parametrization. In terms of the intrinsic Carrollian parameters $(\hat{\sigma},\hat{\chi},\hat{\beta})$, the same transformations need not appear as mutually commuting sectors. In particular, the volume-preserving Carroll--Weyl rescaling acts non-trivially on the normalization of the intrinsic Carroll-boost parameter, as already manifest in \eqref{eq. hatbetaconf}.

As in Einstein gravity, the parameter $h$ remains absent from the transformation of the Carrollian boundary dyad. More importantly, in the conformal theory it drops out entirely from the surface charge, unlike in the Einstein-gravity phase space discussed above. This difference can be traced back to the distinct invariant bilinear forms entering the two Chern--Simons theories. In particular, while Einstein gravity relies on a nonvanishing pairing between Lorentz and translation generators, the conformal bilinear form instead satisfies $\mathrm{Tr}(J_nP_m)=0$ and pairs translations with the special conformal generators $K_n$. It is furthermore convenient to introduce
\begin{equation} \label{eq. PhiUpsilon}
    \Phi := \varphi+\varpi \, , \qquad \Upsilon := \delta_\varepsilon\Phi = \sigma+\chi \, .
\end{equation}
In terms of these variables, the asymptotic surface charge takes the
compact form
\begin{equation} \label{eq. conformal-charge}
    \mathrm{H}_\varepsilon = -\frac{\kappa}{2\pi} \int \mathrm{d}\phi \left[ -2YM +\dot\beta\,\Phi +\Upsilon\left(\Phi'-2\dot B\right) +4\chi\,\varpi' \right] .
\end{equation}
The different terms in \eqref{eq. conformal-charge} demonstrate that the two independent Carroll--Weyl sectors, as well as a non-trivial Carroll-boost sector, admit finite and integrable canonical generators within the present phase space. The combination $\Upsilon=\sigma+\chi$ couples to the current $\Phi'-2\dot B$, while the additional term $4\chi\varpi'$ distinguishes the volume-preserving Carroll--Weyl transformation from its volume-modulating counterpart. In particular, if the two transformations are smeared with the same function, one finds
\begin{equation}
    \mathrm{H}_\chi-\mathrm{H}_\sigma = -\frac{2\kappa}{\pi} \int \mathrm{d}\phi\, \chi\,\varpi' \, ,
\end{equation}
showing that their canonical generators are independent for a generic configuration with non-trivial $\varpi$. Similarly, the term $\dot\beta\,\Phi$ in \eqref{eq. conformal-charge} provides a nonvanishing canonical generator for generic time-dependent Carroll boosts.

It is important, however, to distinguish the algebra of residual gauge transformations from the algebra of non-trivial asymptotic symmetries. Although the residual transformations of the present solution space contain the full $\mathfrak{bms}_3$ sector parametrized by $(Y,T)$, the surface charge \eqref{eq. conformal-charge} is independent of the supertranslation parameter $T$. Supertranslations therefore lie in the kernel of the conformal presymplectic form on the present phase space and are proper gauge transformations rather than non-trivial asymptotic symmetries. The same is true of the radial transformation parametrized by $h$. Likewise, Carroll boosts with~$\dot\beta=0$ carry vanishing charge.

Consequently, the present construction should not be interpreted as a canonical realization of every residual Carrollian transformation. Rather, it provides a minimal asymptotically flat conformal phase space in which both independent Carroll--Weyl rescalings and a non-trivial sector of local Carroll boosts are realized canonically, while the supertranslation and radial sectors remain presymplectic degeneracies. This is sufficient for our present purpose, which is to exhibit the additional canonical Carroll--Weyl symmetry that is obstructed in Einstein gravity.

We also stress that the asymptotic symmetry structure obtained here differs from the nonlinear conformal extension of $\mathfrak{bms}_3$ constructed in \cite{Fuentealba:2020zkf}. The latter arises from a different set of boundary conditions for three-dimensional conformal gravity and contains, in addition to non-trivial supertranslations, an infinite tower of superdilatation and superspecial-conformal generators. By contrast, as emphasized several times above, our boundary conditions are deliberately restricted to the minimal sector required to realize the two Carroll--Weyl transformations and Carroll boosts. In particular, we have not activated an independent $K_n$ sector, which, in view of the non-degenerate pairing
\begin{equation}
    \mathrm{Tr}(P_n K_m)\neq 0 \, ,
\end{equation}
is naturally expected to participate in a larger phase space supporting non-trivial translation charges.

A further difference concerns the boundary dressing fields. In order to realize the Carroll--Weyl transformations canonically, we retain independent fields along both the bulk dilatation generator $D$ and the Lorentz Cartan generator $J_0$, with unrestricted dependence on $(u,\phi)$. This contrasts with the boundary conditions of \cite{Fuentealba:2020zkf}, where the $J_0$ component is not independently excited, while the mode along $D$ has a constrained time dependence determined by the choice of fixed chemical potentials. Since
\begin{equation}
    D=\frac{1}{2}\left(T_+ + T_-\right) ,
\end{equation}
one sees that such a restriction does not retain the two Carroll--Weyl directions independently. Rather, the ``Weyl-type'' transformations present in that phase space are encoded in the superdilatation sector, acting on the celestial circle at null infinity, whereas our boundary conditions are designed to keep the two independent Carroll--Weyl rescalings manifest already at the level of the whole boundary frame.

\subsection{Asymptotic symmetry algebra}

We now determine the algebra of the non-trivial canonical generators of the conformal-gravity phase space. It is important from the outset to distinguish this algebra from the residual gauge algebra discussed above. The latter contains the full $\mathfrak{bms}_3$ sector parametrized by $(Y,T)$, together with the additional residual transformations $(\sigma,\chi,\beta,h)$. Some of these transformations, however, lie in the kernel of the conformal presymplectic form and therefore have vanishing canonical generators.

Using the Poisson bracket~\eqref{eq. PB} and the double-mode basis $F_{pq}$ \eqref{eq. doublemodebasis} introduced in the previous Section, we define
\begin{equation}
    \begin{aligned}
    &\mathcal{J}_n := \mathrm{H}_\varepsilon\left(Y\sim\mathrm{e}^{\mathrm{i}n\phi}\right), &&\mathcal{W}^{+}_{pq} := \mathrm{H}_\varepsilon\left(\sigma \sim F_{pq}\right),\\
    &\mathcal{W}^{-}_{pq} := \mathrm{H}_\varepsilon\left(\chi \sim F_{pq}\right), &&\mathcal{B}_{pq} := \mathrm{H}_\varepsilon\left(\beta \sim F_{pq}\right).
    \end{aligned}
\end{equation}
The supertranslation and radial modes vanish identically,
\begin{equation}
    \mathcal{P}_n := \mathrm{H}_\varepsilon\left(T\sim\mathrm{e}^{\mathrm{i}n\phi}\right) =0 \, , \qquad \mathcal{R}_{pq} := \mathrm{i}\mathrm{H}_\varepsilon\left(h \sim F_{pq}\right) =0 \, .
\end{equation}
They therefore belong to the kernel of the conformal presymplectic form for the phase space considered here. In particular, although the residual gauge transformations contain the full $\mathfrak{bms}_3$ sector, its canonical realization does not: supertranslations are proper gauge transformations in the present symplectic prescription. After quotienting by these degenerate directions, the non-trivial gravitational diffeomorphism sector is therefore reduced to the superrotation sector.

The complete set of nonvanishing Poisson brackets among the remaining canonical generators is
\begin{subequations} \label{eq. full-conformal-Carroll-algebra}
    \begin{align}
        \mathrm{i} \left\{ \mathcal{J}_n,\mathcal{J}_m \right\} &= (n-m)\mathcal{J}_{n+m} -\frac{c_1}{12} n^3\delta_{n+m,0} \, ,\\
        \mathrm{i} \left\{ \mathcal{W}^{+}_{pq},\mathcal{W}^{+}_{rs} \right\} &= -\frac{c_1}{12} (p-q) \mathrm{e}^{2\mathrm{i}(q+s)u} \delta_{p+r,q+s} \, ,\\
        \mathrm{i} \left\{ \mathcal{W}^{+}_{pq},\mathcal{W}^{-}_{rs} \right\} &= -\frac{c_1}{12} (p-q) \mathrm{e}^{2\mathrm{i}(q+s)u} \delta_{p+r,q+s} \, ,\\
        \mathrm{i} \left\{ \mathcal{W}^{-}_{pq},\mathcal{W}^{-}_{rs} \right\} &= -\frac{5c_1}{12} (p-q) \mathrm{e}^{2\mathrm{i}(q+s)u} \delta_{p+r,q+s} \, ,\\
        \mathrm{i} \left\{ \mathcal{W}^{+}_{pq},\mathcal{B}_{rs} \right\} &= -\frac{c_1}{12} (r+s) \mathrm{e}^{2\mathrm{i}(q+s)u} \delta_{p+r,q+s} \, ,\\
        \mathrm{i} \left\{ \mathcal{W}^{-}_{pq},\mathcal{B}_{rs} \right\} &= -\frac{c_1}{12} (r+s) \mathrm{e}^{2\mathrm{i}(q+s)u} \delta_{p+r,q+s} \, .
    \end{align}
\end{subequations}
Here,
\begin{equation}
    c_1=12\kappa \, .
\end{equation}
All remaining brackets vanish in the compensated parametrization used here. In particular,
\begin{equation} \label{eq. boostnoselfextension}
    \left\{ \mathcal{B}_{pq},\mathcal{B}_{rs} \right\} =0 \, .
\end{equation}

After quotienting by the presymplectic kernel, the gravitational diffeomorphism sector is therefore chiral and consists of a single centrally extended Virasoro algebra. In the conventional notation for the two central terms of $\mathfrak{bms}_3$, conformal gravity corresponds to~$c_1=12\kappa$ and $c_2=0$, in contrast with three-dimensional Einstein gravity, where $c_1=0$ and $c_2=12\kappa$. In the present phase space the statement is in fact stronger than $c_2=0$: the supertranslation generators $\mathcal{P}_n$ themselves vanish.

This difference has a direct Chern--Simons origin, as already noted at the level of the charges; we briefly restate the argument here for clarity. In Einstein gravity the invariant bilinear form pairs Lorentz transformations and translations, $\mathrm{Tr}(J_nP_m)\neq0$, whereas the conformal invariant bilinear form obeys $\mathrm{Tr}(J_nP_m)=0$ and $\mathrm{Tr}(P_nK_m)\neq0$. Since the minimal phase space constructed here does not activate an independent $K_n$ sector, the translational directions have no canonically paired boundary data and remain presymplectic degeneracies. Recovering non-trivial supertranslation charges would therefore require an enlargement of the present phase space. This is consistent with the fact that larger boundary conditions for three-dimensional conformal gravity lead to the conformal extension of $\mathfrak{bms}_3$, where additional conformal sectors are retained as non-trivial boundary symmetries \cite{Fuentealba:2020zkf,Fuentealba:2024thk}.

The two Carroll--Weyl currents, on the other hand, form a rank-two centrally extended Abelian current subalgebra in \eqref{eq. full-conformal-Carroll-algebra}. Their relative levels are encoded in the matrix
\begin{equation} \label{eq. relative-coef}
    K_\mathrm{CW} =
    \begin{pmatrix}
        1 & 1 \\
        1 & 5
    \end{pmatrix}.
\end{equation}
Since the determinant of this matrix is non-zero, the two local Carroll--Weyl current directions are canonically independent. This statement concerns the two current species; as usual for an affine cocycle involving a spatial derivative, spatially constant zero modes may lie in the kernel of the central extension. The Carroll boosts have no self-extension \eqref{eq. boostnoselfextension} but participate in equal mixed central extensions with the two Carroll--Weyl currents. This should be contrasted with Einstein gravity, where the single Carroll--Weyl symmetry forms a canonical pair with its radial partner and the Carroll boosts possess a nonvanishing boost--boost central extension.

Interestingly, the relative coefficients in \eqref{eq. relative-coef} admit a simple algebraic interpretation. Recall that the two physical Carroll--Weyl transformations are generated by $T_+ = J_0$ and $T_- = 2D - J_0$ \eqref{eq. TpTm}. Using the invariant bilinear form of $\mathfrak{so}(3,2)$, one immediately obtains
\begin{equation}
    \mathrm{Tr}(T_+T_+)=1 \, , \qquad \mathrm{Tr}\!\left(T_+(-T_-)\right)=1 \, , \qquad \mathrm{Tr}\!\left((-T_-)(-T_-)\right)=5 \, .
\end{equation}
Thus, up to the common overall Chern--Simons level, $K_{\mathrm{CW}}$ is precisely the Gram matrix of the two physical Carroll--Weyl generators with respect to the invariant bilinear form. The non-diagonal form of this matrix is therefore simply a consequence of using the physically natural Carroll--Weyl basis $(T_+,-T_-)$ rather than an orthogonal basis of the Cartan subalgebra. Indeed, introducing
\begin{equation}
    T_D := -T_- -T_+ = -2D \, ,
\end{equation}
one finds
\begin{equation}
    \mathrm{Tr}(T_+T_D)=0 \, , \qquad \mathrm{Tr}(T_DT_D)=4 \, ,
\end{equation}
so that in the equivalent basis $(T_+,T_D)$ the level matrix becomes
\begin{equation}
    K_{\mathrm{CW}} \longrightarrow
    \begin{pmatrix}
        1 & 0\\
        0 & 4
    \end{pmatrix}.
\end{equation}
The coefficients $(1,1,5)$ appearing in the asymptotic charge algebra therefore follow directly from the embedding of the two physical Carroll--Weyl directions into the Cartan subalgebra of $\mathfrak{so}(3,2)$. Their non-degenerate Gram matrix gives an algebraic and basis-independent characterization of the canonical independence of the two Carroll--Weyl currents.

This observation further clarifies the relation with the boundary conditions of~\cite{Fuentealba:2020zkf}. Since their phase space does not contain an independent component along $J_0$, only the dilatation direction $D$ is activated within the Cartan subalgebra. In terms of the physical Carroll--Weyl generators, however, $D=\frac{1}{2}\left(T_+ + T_-\right)$, so that retaining the $D$ direction alone selects only a particular linear combination of the two Carroll--Weyl transformations. The absence of an independent $J_0$ component therefore prevents the two Carroll--Weyl directions from being resolved as independent canonical currents. Equivalently, the characteristic mixing between $D$ and $J_0$ that leads, in the physical $(T_+,-T_-)$ basis, to the non-diagonal Gram matrix above is absent in that phase space. This provides a simple algebraic explanation for why the two independent Carroll--Weyl symmetries could not have been seen separately within the boundary conditions of \cite{Fuentealba:2020zkf}.

\subsection{Variational principle and anomalies}

We finally apply the variational analysis of Section~\ref{sec. grav} to the conformal-gravity phase space. With the same boundary term~\eqref{eq. CHvD}, evaluated using the conformal connection \eqref{eq. conf-aCBW} and bilinear form~\eqref{eq. so32-bilinear}, the on-shell contribution at future null infinity reads
\begin{equation} \label{eq. conf-onshell-variation}
    \left.\delta (S+S^{(0)}_\mathrm{bdy})\right|_{\mathrm{EOM}} = \frac{\kappa}{2\pi} \int \mathrm{d}u\mathrm{d}\phi \left[ 2\ddot B\,\delta\Phi -2\dot\Phi\,\delta\dot B -\dot\Phi'\,\delta\Phi -4\dot\varpi'\,\delta\varpi \right] \equiv \int\mathrm{d}u\mathrm{d}\phi\,\Theta \, ,
\end{equation}
where we have used the combinations~\eqref{eq. PhiUpsilon} and have integrated by parts along the periodic celestial circle. Total retarded-time derivatives are instead retained, since they contribute at the initial and final cuts of~$\mathscr I^+$. As in Einstein gravity, the remaining local boundary one-form is not exact in field space. Its curvature is
\begin{equation}
    \delta\Theta = \frac{\kappa}{2\pi} \Big[ 2\,\delta\ddot B\wedge\delta\Phi -2\,\delta\dot\Phi\wedge\delta\dot B -\delta\dot\Phi'\wedge\delta\Phi -4\,\delta\dot\varpi'\wedge\delta\varpi \Big] \, ,
\end{equation}
which does not vanish for unrestricted variations of the boundary fields. Thus, a boundary counterterm alone cannot cancel this local variational contribution while leaving the corner prescription unchanged. For field-independent residual labels $\bar{\varepsilon}=(\sigma,\chi,\beta)$, the transformations $\delta_{\bar{\varepsilon}}\Phi=\Upsilon$, $\delta_{\bar{\varepsilon}}\varpi=\chi$ and $\delta_{\bar{\varepsilon}}B=\beta/2$ give
\begin{equation} \label{eq. conf-anomalous-variation}
    \mathcal A[\bar{\varepsilon}] := \delta_{\bar{\varepsilon}} S = \frac{\kappa}{2\pi} \int\mathrm{d}u\,\mathrm{d}\phi \left[ -\Upsilon\dot\Phi' -4\chi\dot\varpi' +2\Upsilon\ddot B -\dot{\beta}\dot\Phi \right].
\end{equation}
This is the anomalous contribution from~$\mathscr I^+$ to $\delta_{\bar{\varepsilon}}(S+S^{(0)}_{\mathrm{bdy}})$ in the present symplectic prescription.

The corner ambiguity discussed in the Einstein-gravity analysis again allows this boundary variational contribution to be removed \cite{Jacobson:1993vj,Iyer:1994ys,Freidel:2020xyx,Ciambelli:2023ott,Ciambelli:2024vhy,Delfante:2025lxn}. A suitable combined improvement is
\begin{subequations} \label{eq. conf-full-improvement}
    \begin{align}
        \frac{4\pi}{\kappa}\vartheta_{\mathrm{corner}} &= \left(\Phi'-4\dot B\right)\delta\Phi +4\varpi'\delta\varpi \, ,\\
        \frac{4\pi}{\kappa}\ell_{\mathrm{bdy}} &= 4\dot\Phi\,\dot B -\Phi'\dot\Phi -4\varpi'\dot\varpi \, .
    \end{align}
\end{subequations}
Indeed, one finds the local identity
\begin{equation} \label{eq. conf-improved-variation}
    \Theta +\partial_u\vartheta_{\mathrm{corner}} +\delta\ell_{\mathrm{bdy}} = -\frac{\kappa}{4\pi} \partial_\phi
    \left( \dot\Phi\,\delta\Phi +4\dot\varpi\,\delta\varpi \right) ,
\end{equation}
whose right-hand side integrates to zero around the celestial circle. Adding $\ell_{\mathrm{bdy}}$ alone therefore leaves a temporal corner contribution; the combined prescription also cancels the boundary variational potential. The effect on the canonical charges is controlled by
\begin{equation} \label{eq. conf-corner-curvature}
    \delta\vartheta_{\mathrm{corner}} = \frac{\kappa}{2\pi} \left[ -2\delta\dot B\wedge\delta\Phi +\frac12\delta\Phi'\wedge\delta\Phi +2\delta\varpi'\wedge\delta\varpi \right].
\end{equation}
For fixed residual labels, its contribution to the surface-charge variation is
\begin{equation}
    \Delta(\delta\mathrm H_\varepsilon) = \int \mathrm{d}\phi\, \delta\vartheta_{\mathrm{corner}} (\delta,\delta_\varepsilon) = -\delta\mathrm H_\varepsilon^{\mathrm{Carroll}} \, ,
\end{equation}
where $\mathrm H_\varepsilon^{\mathrm{Carroll}}$ denotes the $(\sigma,\chi,\beta)$ contribution to \eqref{eq. conformal-charge}. With the same reference normalization, the improved charge consequently becomes
\begin{equation}
    \mathrm H_\varepsilon^{\mathrm{improved}} = \frac{\kappa}{\pi} \int \mathrm{d}\phi\,YM \, .
\end{equation}
Both Carroll--Weyl sectors and the Carroll boosts therefore become canonically trivial under this improvement. The non-trivial realization established above instead uses the original corner prescription, together with the boundary variation~\eqref{eq. conf-anomalous-variation}.

The relation to the central extensions follows from the same symplectic descent as in~\eqref{eq. anomaly-descent}. Since the transformations labeled by $\bar{\varepsilon}$ commute, the antisymmetrized variation satisfies
\begin{equation}
    \delta_{\bar{\varepsilon}_1}\mathcal A[\bar{\varepsilon}_2] - \delta_{\bar{\varepsilon}_2}\mathcal A[\bar{\varepsilon}_1] = - \int\mathrm{d}u\,\partial_u K(\bar{\varepsilon}_1,\bar{\varepsilon}_2) \, .
\end{equation}
Writing $\lambda_i^I=(\sigma_i,\chi_i)$ and using the matrix $K_{\mathrm{CW}}$ defined in~\eqref{eq. relative-coef}, the representative compatible with the canonical charges is
\begin{align} \label{eq. conf-anomaly-cocycle}
    K(\bar{\varepsilon}_1,\bar{\varepsilon}_2) &= - \frac{\kappa}{4\pi} \int\mathrm{d}\phi \left[ \lambda_1^I(K_{\mathrm{CW}})_{IJ} \partial_\phi\lambda_2^J - \lambda_2^I(K_{\mathrm{CW}})_{IJ} \partial_\phi\lambda_1^J \right] \nonumber\\
    &\quad +\frac{\kappa}{2\pi} \int\mathrm{d}\phi \left[ \Upsilon_1\dot{\beta}_2 - \Upsilon_2\dot{\beta}_1 \right].
\end{align}
In particular, with the Poisson-bracket convention~\eqref{eq. PB}, one verifies that $K(\bar{\varepsilon}_1,\bar{\varepsilon}_2) =\delta_{\bar{\varepsilon}_2}\mathrm H_{\bar{\varepsilon}_1}$. It therefore reproduces precisely the Carrollian central extensions in~\eqref{eq. full-conformal-Carroll-algebra}: the Carroll--Weyl levels have relative coefficients $(1,1,5)$, the two mixed Weyl--boost extensions coincide, and no boost--boost cocycle occurs. As in Einstein gravity, the Wess--Zumino consistency condition thus holds modulo temporal corner contributions, whose descent yields the canonical two-cocycle.

The new feature relevant to the possible comparison with Carroll--Weyl-gauged tensionless strings~\cite{Sheikh-Jabbari:2026vqh,Duary:2026rlo} is therefore the rank-two Weyl level matrix. Its entries are fixed by the bulk invariant form and the single coupling~$\kappa$; they are not independent quantum anomaly coefficients. Relating this classical structure to the distinct worldsheet cocycles~\cite{Duary:2026lmk,Chen:2026cau} would require a map between the relevant symmetry algebras and a matching of their cohomology classes and normalizations. Whether such a map can arise through holographic anomaly matching or Chern--Simons anomaly inflow~\cite{Callan:1984sa} remains an open question.


\section{Conclusion} \label{sec. Conclu}

In this work, we have investigated whether the volume-preserving Carroll--Weyl rescaling can be realized as a genuine asymptotic symmetry at future null infinity of a three-dimensional asymptotically flat bulk. This transformation rescales the temporal and spatial components of the Carrollian coframe with opposite weights and has no counterpart as an independent Weyl rescaling of a non-degenerate Lorentzian metric. Its intrinsic geometric existence does not, however, guarantee a holographic realization: it must first arise from an admissible residual bulk gauge transformation and then possess a finite, integrable and generically nonvanishing canonical generator. Our analysis establishes both the obstruction in Einstein gravity and a realization in three-dimensional conformal gravity.

In Einstein gravity, the covariant Bondi--Weyl phase space \eqref{eq. aCBW} accommodates charged superrotations, supertranslations, the volume-modulating Carroll--Weyl transformation and local Carroll boosts. The volume-preserving rescaling encounters an obstruction. Geometrically, the standard conformal completion correlates the rescalings of the induced degenerate metric and its kernel vector, in a way that it excludes that the respective Weyl weights be opposite. Algebraically, the commuting internal directions available alongside the common rescaling generated by $J_0$ satisfy $\operatorname{Cent}_{\mathfrak{iso}(1,2)}(J_0) = \operatorname{span}\{J_0,P_0\}$ and $(\operatorname{ad}_{P_0})^2=0$. Modulo $J_0$, the only candidate is therefore the nilpotent translation $P_0$. On the leading Bondi frame, $J_0$ produces a common diagonal rescaling, while $P_0$ produces the triangular action of a Carroll boost; neither supplies a second diagonal action with opposite weights. Under the assumptions of an invertible bulk triad and the standard $\mathrm{ISO}(1,2)$ Chern--Simons description, the obstruction therefore precedes the construction of surface charges: modulo boundary diffeomorphisms, the second Carroll--Weyl transformation is not generated as an additional independent internal residual symmetry. Enlarging the boundary data may expose a corresponding kinematical direction in solution space, but it cannot by itself supply the missing independent semisimple gauge direction required for its residual realization. The obstruction is therefore structural, rather than a consequence of a vanishing charge in the particular phase space considered here.

Conformal gravity removes this obstruction by rendering local Weyl transformations of the physical bulk metric genuine gauge symmetries. Actually, in its $\mathfrak{so}(3,2)$ Chern--Simons formulation, bulk dilatations provide the additional commuting semisimple direction. The combinations $T_+=J_0$ and $T_-=2D-J_0$ generate the volume-modulating and volume-preserving rescalings, respectively. Allowing the associated bulk fields $\varphi(u,\phi)$ and $\varpi(u,\phi)$ to fluctuate independently gives a residual realization of both Weyl transformations, together with local Carroll boosts, and hence of the complete internal symmetry sector of the Carrollian frame. We further establish their canonical realization: both Weyl sectors and generic time-dependent boosts possess finite, integrable and nonvanishing surface generators~\eqref{eq. conformal-charge}. This statement concerns the canonical realization of the internal frame-symmetry sectors and does not imply that every residual transformation carries a non-trivial charge. In particular, boosts whose adapted parameter obeys $\dot\beta=0$, as well as the entire supertranslation sector of residual boundary diffeomorphisms, lie in the kernel of the presymplectic form on the phase space considered here and therefore remain proper gauge transformations.

The charge algebra \eqref{eq. full-conformal-Carroll-algebra} makes the independence of the two Carroll--Weyl currents explicit. Up to the common Chern--Simons level $\kappa$, their level matrix is the Gram matrix of~$(T_+,T_-)$ under the invariant bulk bilinear form \eqref{eq. relative-coef}. Its non-degeneracy establishes the canonical independence of the two local current species. The non-trivial diffeomorphism sector consists of a single Virasoro algebra with classical central charge $c_1=12\kappa$, while $c_2=0$ in the usual~$\mathfrak{bms}_3$ notation. In the compensated parametrization, the boost current has no self-extension but enters mixed central extensions with both Weyl currents. This differs substantially from Einstein gravity, where the boost sector has its own nonvanishing cocycle and the single Weyl sector pairs with the radial mode. The change reflects the different invariant bilinear forms, and therefore the different symplectic structures, of the two bulk theories.

The variational analysis connects these canonical structures to the boundary anomaly. The antisymmetrized variation of the anomalous boundary term descends to the Carroll--Weyl and mixed boost cocycles \eqref{eq. conf-anomaly-cocycle}. A corner improvement \eqref{eq. conf-full-improvement}, accompanied by suitable boundary terms, removes the anomalous variational contribution, but simultaneously renders the corresponding frame transformations canonically trivial. Thus the boundary prescription determines both the invariance properties of the variational potential and which frame degrees of freedom remain charged. The anomaly, the corner ambiguity and the central extensions are linked by the same symplectic descent. These are classical statements about the gravitational phase space; identifying a quantum boundary theory and its anomalies requires further input.

Our construction complements earlier boundary conditions of conformal gravity with charged Weyl factors or several Cartan currents~\cite{Afshar:2013bla,Lovrekovic:2023xsj}. Its contribution is the intrinsic identification and simultaneous canonical realization of the two Carroll--Weyl rescalings at null infinity. It also offers a geometric perspective on the bulk dilatation direction underlying the superdilatations of conformal $\mathfrak{bms}_3$~\cite{Fuentealba:2020zkf} and its enhanced analogue in extended conformal gravity~\cite{Fuentealba:2024thk}. In fact, we retain independent boundary dressing fields along both $D$ and the Lorentz Cartan generator $J_0$, with arbitrary dependence on $(u,\phi)$. The distinction from those reductions \cite{Afshar:2013bla,Fuentealba:2020zkf,Lovrekovic:2023xsj,Fuentealba:2024thk} concerns this freedom: their time dependence is constrained by fixed chemical potentials or chiral evolution, whereas our frame transformations are local on the full Carrollian boundary, with charges evaluated on its celestial cuts. Since $D=\tfrac12(T_++T_-)$, bulk dilatations actually enter a definite combination of the two intrinsic Weyl actions. This places the dilatation direction within a broader local frame description, without identifying the different reduced phase spaces or their charge algebras.

Several directions follow from these results. First, one should determine whether a broader asymptotically flat phase space can retain the present Weyl and boost sectors while restoring non-trivial supertranslation charges. Supertranslations survive residually in our construction but have vanishing canonical generators, unlike in Einstein gravity; this distinction is already visible in the flat conformal-gravity analysis of~\cite{Afshar:2013bla}. The mechanism is transparent: translations pair with special conformal generators through $\mathrm{Tr}(P_nK_m)$, while our minimal phase space contains no independent $K_n$ data. The conformal $\mathfrak{bms}_3$ reduction of~\cite{Fuentealba:2020zkf} retains such data and realizes charged supertranslations together with superdilatations and superspecial conformal transformations, with a related enhancement in~\cite{Fuentealba:2024thk}. Whether these sectors can coexist with both local Carroll--Weyl currents and the boost sector, while preserving the finiteness and integrability of the charges, is left for future work. The present work provides a minimal asymptotically flat phase space of three-dimensional conformal gravity in which both Carroll--Weyl rescalings and local Carroll boosts are realized as residual symmetries with generically nonvanishing canonical generators. Extending this construction also calls for an intrinsic boundary interpretation of the additional conformal directions, beyond their identification with bulk generators. Tracking these directions between asymptotically flat and asymptotically AdS boundary conditions should further clarify how their geometric interpretation changes when the boundary becomes timelike and admits only one ordinary metric Weyl rescaling.

Second, the relation between the corner improvement \eqref{eq. conf-full-improvement} and the disappearance of the frame charges motivates an extended phase-space treatment. Edge-mode constructions restore gauge invariance by introducing boundary variables while allowing a non-trivial algebra of surface symmetries~\cite{Donnelly:2016auv,Geiller:2017whh}. Applied here, they could determine whether these two boundary prescriptions admit a common gauge-invariant extension in which the Carroll--Weyl and boost currents act on explicit edge degrees of freedom. Such a construction would clarify the distinction between compensating bulk gauge transformations and the physical boundary symmetries carrying the current algebra.

Third, an explicit two-dimensional boundary theory for our phase space should be derived. The Chern--Simons/Wess--Zumino--Witten correspondence and its Hamiltonian reductions provide a natural starting point~\cite{Wess:1971yu,Witten:1983ar,Elitzur:1989nr,Coussaert:1995zp,Barnich:2013yka}. The reduction appropriate to the present boundary conditions should determine how the two diagonal fields couple to the boost sector and reproduce the full current algebra, including its mixed cocycles. It would also provide access to boundary sources, Ward identities, representations and quantization, including possible quantum corrections to the classical levels.

Such a construction would also provide a concrete setting in which to explore a dynamical relation with Carroll--Weyl-gauged null strings. The structural parallel discussed in the main text offers a starting point: realizing the volume-preserving rescaling as an unrestricted local symmetry requires an additional Carroll--Weyl connection in the gauged worldsheet theory~\cite{Sheikh-Jabbari:2026vqh,Sheikh-Jabbari:2026tpf}, while our gravitational construction requires an independent dilatation direction in the bulk gauge algebra. The relation of the gauged model to the conformal null string in Dirac space~\cite{Lindstrom:2026zno} further motivates investigating whether this parallel extends to the respective actions and symplectic structures. A first step would be to look for counterparts of the worldsheet Carroll--Weyl connection and scaling constraint in the reduced boundary theory, accounting for the distinction between worldsheet gauge redundancies and charged surface symmetries. Together with a boundary quantization, such a map would provide a basis for comparing the gravitational cocycles with the quantum Carroll--Weyl and BRST anomalies of null strings~\cite{Duary:2026rlo,Duary:2026lmk,Chen:2026cau}. Matching the relevant cohomology classes and their coefficients would then provide a concrete test of a possible holographic anomaly-matching or anomaly-inflow interpretation~\cite{Callan:1984sa}.

Fourth and finally, the relation between current levels and the invariant bulk bilinear form motivates an extension to conformal higher-spin gravity~\cite{Pope:1989vj,Grigoriev:2019xmp,Lovrekovic:2023xsj}. Additional Cartan directions could act on an enlarged set of boundary geometric fields and support further independent currents, extending the boundary freedom explored in higher-spin gravity~\cite{Delfante:2025lxn}. Their interpretation must be established intrinsically: increasing the rank of the bulk algebra does not by itself create additional Weyl rescalings of an ordinary two-dimensional Carrollian coframe. The challenge is therefore to identify how these directions act on higher-spin boundary data and which acquire non-trivial canonical generators under suitable boundary conditions. The compatibility of this enlarged boundary freedom with finite, integrable charges and a well-defined variational principle should be examined in parallel. Comparison with existing analyses of action and charge renormalization~\cite{Campoleoni:2023eqp,Campoleoni:2025bhn} could clarify the role of boundary and corner prescriptions in selecting admissible higher-spin phase spaces. A complementary direction concerns asymptotically FLRW spacetimes~\cite{Campoleoni:2026abr}, whose cosmological scale factor motivates exploring analogous boundary conditions in conformal gravity and investigating how cosmological expansion interacts with bulk Weyl freedom, the boundary Carroll--Weyl currents and the finiteness of their charges. These developments would extend the present relation between bulk gauge symmetry, intrinsic boundary geometry and canonical current algebras.

The broader lesson is that the degeneracy of null infinity should not be viewed merely as an obstacle to importing familiar Lorentzian notions of conformal symmetry. It creates an additional intrinsic rescaling that asks for a larger bulk gauge completion. In three dimensions, conformal gravity supplies that completion: the missing volume-preserving Carroll--Weyl transformation becomes an independent charged boundary current, and both intrinsic Weyl rescalings acquire a simultaneous canonical realization at future null infinity.


\section*{Acknowledgments}

We would like to thank Euihun Joung for useful discussions. The work of A.D. was supported by the Brain Pool Program funded by the Ministry of Science and ICT through the National Research Foundation of Korea (RS-2025-25457100). The work of C.M. was supported by a fellowship of the Scuola Normale Superiore, the ERC (NOTIMEFORCOSMO, 101126304) and the INFN (I.S. GSS-Pi). A.D. thanks the Asia Pacific Center for Theoretical Physics (APCTP) in Pohang for hospitality during the completion of this work, as well as the organizers and participants of the workshop ``Developments in QFT, Gravity, and Holography'' held at APCTP, where preliminary results were presented. C.M. thanks the Aristotle University of Thessaloniki for hospitality during the workshop ``Carroll Think Tank: 3rd edition,'' where part of these results were presented.  


\bibliographystyle{JHEP}

\providecommand{\href}[2]{#2}
\end{document}